\documentclass{aa}
\usepackage{graphicx,amssymb}
\usepackage{array}
\usepackage{natbib}
\usepackage{caption}
\usepackage{amsmath}
\usepackage{bm}
\begin{document}

\title{Clustering of the AKARI NEP Deep Field 24 $\mu$m selected galaxies}

\author{A.~Solarz\inst{1,2}
  \and A.~Pollo\inst{2,3}
\and T.~T.~Takeuchi\inst{1}
\and K.~Ma\l ek\inst{1,2}
\and H.~Matsuhara\inst{4}
\and G. J.~White\inst{5,6}
\and A.~P\c epiak\inst{2}
\and T.~Goto\inst {7,8}
\and T.~Wada\inst{4}
\and S.~Oyabu\inst{1}
\and T.~Takagi\inst{4}
\and Y.~Ohyama\inst{9}
\and C.~P.~Pearson\inst{5,6,10}
\and H.~Hanami\inst{11}
\and T.~Ishigaki\inst{11}
\and M.~Malkan\inst{12}
}

\institute{Division of Particle and Astrophysical Science, Nagoya University, Furo-cho, Chikusa-ku, Nagoya 464-8602, Japan\\
         \email{aleksandra.solarz@ncbj.gov.pl}
         \and
         National Center for Nuclear Research, ul.\ Ho\.{z}a 69, 00-681 Warsaw, Poland 
         \and      
         The Astronomical Observatory of the Jagiellonian University, ul.\ Orla 171, 30-244 Krak\'{o}w, Poland
        \and
         Institute of Space and Astronautical Science, Japan Aerospace Exploration Agency, Sagamihara, Kanagawa 252-5210, Japan
         \and
         Department of Physics and Astronomy, The Open University, Walton Hall, Milton Keynes, MK7 6AA, UK
         \and
        Space Science and Technology Department, CCLRC Rutherford Appleton Laboratory, Chilton, Didcot, Oxfordshire OX11 0QX, UK
        \and
        Institute for Astronomy, University of Hawaii, 2680 Woodlawn Drive, Honolulu, HI, 96822, USA
        \and
        National Astronomical Observatory, 2-21-1 Osawa, Mitaka, Tokyo, 181-8588, Japan
        \and
        Academia Sinica, Institute of Astronomy and Astrophysics, Taiwan
        \and
        Department of Physics, University of Lethbridge, 4401 University Drive, Lethbridge, Alberta T1J 1B1, Canada
        \and
        Physics Section, Faculty of Humanities and Social Sciences, Iwate University, Morioka 020-8550, Japan
         \and
         Departament of Physics and Astronomy, University of California, Los Angeles, CA 90024, USA
        }

\date{Received <date>/ Accepted <date>}

\abstract{}
{We present a method of selection of 24~$\mu$m galaxies from the AKARI North Ecliptic Pole (NEP) Deep Field down to $150 \mbox{ }\mu$Jy and measurements of their two-point correlation function.
We aim to associate various 24 $\mu$m selected galaxy populations with present day galaxies and to investigate the impact of their environment on the direction of their subsequent evolution.}
{We discuss using  of Support Vector Machines (SVM) algorithm applied to infrared photometric data to perform star-galaxy separation, in which we achieve an accuracy higher than 80\%.
The photometric redshift information, obtained through the CIGALE code, is used to explore the redshift dependence of the correlation function parameter ($r_{0}$) as well as the linear bias evolution. This parameter relates galaxy distribution to the one of the underlying dark matter.
We connect the investigated sources to their potential local descendants through a simplified model of the clustering evolution without interactions.}
{We observe two different populations of star-forming galaxies, at $z_{med}\sim 0.25$, $z_{med}\sim 0.9$. Measurements of total infrared luminosities ($L_{TIR}$) show that the sample at $z_{med}\sim 0.25$ is composed mostly of local star-forming galaxies, while the sample at $z_{med}\sim0.9$ is composed of luminous infrared galaxies (LIRGs) with $L_{TIR}\sim 10^{11.62}L_{\odot}$. 
We find that dark halo mass is not necessarily correlated  with the $L_{TIR}$: for subsamples with $L_{TIR}= 10^{11.15} L_{\odot}$ at $z_{med}\sim 0.7$ we observe a higher clustering length ($r_{0}=6.21\pm0.78$ $[h^{-1} \mbox{Mpc}]$) than for a subsample with mean $L_{TIR}=10^{11.84} L_{\odot}$ at $z_{med}\sim1.1$ ($r_{0}=5.86\pm0.69$  $h^{-1} \mbox{Mpc}$).
We find that galaxies at $z_{med}\sim 0.9$ can be ancestors of present day $L_{*}$ early type galaxies, which exhibit a very high $r_{0}\sim 8$~$h^{-1} \mbox{Mpc}$.}
{}
\keywords{infrared: galaxies --
  infrared: stars -- galaxies: fundamental parameters -- galaxies: statistics}

\titlerunning{Clustering of the AKARI NEP 24 $\mu$m galaxies}
\maketitle 

\section{Introduction}

%Observations of the Cosmic Microwave Background Radiation (CMB; e.g. \citealt{bennett13}) have shown a very uniform matter distribution, with only small embedded fluctuations, which gave rise to the vast variety of structures observed in the Universe today.
%Present day galaxies differ in many aspects, like age, stellar content, morphology, spectral type, luminosity and colour. Each property encodes information about the galaxy's formation and its evolution, which means that each one of them is related to their local host environment.
%Understanding the complexity of the structures in the Universe is one of the most enigmatic questions that current observational cosmology faces today.
%At present, the most widely accepted paradigm of structure formation assumes that the galaxies have formed and evolved inside dark matter halos, which grew through accretion and mergers (\citealt{wr}).
The connection between the properties of galaxies, such as morphology, luminosity, colour, surface brightness, or specific star formation rate (SFR), and the local environment in which they reside is well documented (e.g., \citealt{marinoni99}; \citealt{marinoni02}; \citealt{hogg04}; \citealt{weinmann06}; \citealt{hirschmann}; \citealt{boselli}; \citealt{guglielmo}) over a wide range of redshifts. 
Different environments, from rich clusters to very low density areas, can have a significant influence on formation and a subsequent evolution of galaxies. 
A range of environmental mechanisms, such as galaxy mergers (\citealt{toomre}), harrassment (\citealt{moore}), and gas stripping (\citealt{gunn}) are expected to  influence the determination of different galaxy properties. However, intrinsic physical processes, such as supernovae feedback (e.g., \citealt{samuel}) or a central black hole (e.g., \citealt{choi}), could be essential in putting a galaxy into a specific evolutionary path.
One main question remains unanswered: are the observed dependencies a product of a cumulation of many processes over the cosmic time, or were they predetermined at the time of first galaxy assembly.
%nature vs nurture
One of the approaches that can address these issues involves studying the statistical correlation between the fluctuations of mass (\citealt{kaiser}; \citealt{mowhite}; \citealt{bardeen}).
%Evolutionary scenario in which star formation, gas depletion processes are accelerated
%in more  luminous  objects  and  in high  density environments:  star  formation activity is progressively  shifting with cosmic  time towards  lower luminosity galaxies (downsizing), and out of high density environments
%Each property encodes information about the galaxy's formation and its evolution, which means that each one of them is related to their local host environment.
%Therefore, different properties of the galaxies and their host dark matter halos must be intertwined, and it is crucial to provide observational links between them.}
%Since the properties of the galaxies, and the halos inside which they have formed are intertwined,  
%The most successful method of performing this task is based on the clustering estimates. According to the biased hierarchical formation scenario, dark matter itself clusters together (\citealt{mowhite}, \citealt{gao}) with a strength dependent on its mass (\citealt{bardeen}).
In this context, one way to shed light on the nature of the link between the baryonic component and the underlying matter distribution is to explore the properties of a specific type of galaxy, and relate them to the parent dark matter halo (DMH).
%Near infrared (NIR) wavelengths are perfectly suited for tracing the older stellar populations, allowing for the predominant selection of galaxies, which already have formed majority of their stars and are passively evolving.
%Mid infrared (MIR) observations, on the other hand, trace the dust emission resulting from the heating by high-energy photons, and therefore are invaluable in detecting sources actively forming stars.
%Extensive research was performed probing the NIR passbands from Spitzer, 
%Observations of Large Scale Structure in the infrared (IR) wavelengths bring the most sought after constrains on the models of the formation and evolution of galaxies. The 

%{\bf
%In the regime of optical observations of the Universe up to $z\sim 1$ it is a well known fact that the most massive, evolved galaxies are much more strongly clustered than the less massive, actively forming stars ones (e.g. \citealt{zehavi11}, \citealt{meneux06}, \citealt{pollo06}, \citealt{coil06}). In the infrared (IR) part of the spectrum those relations are investigated in a lesser extent (e.g. \citep{torre}, \citep{maglio}, \citep{starikova}).

%, which does not yet allow for a consistent 
%}

As the hierarchical model of structure formation (e.g., \citealt{granato}) predicts, it is expected that galaxy formation, as well as star formation, is intensively driven by mergers. In this case, the first bursts of star formation would occur in the denser regions of the Universe before appearing in the less dense and therefore less clustered regions (\citealt{elbaz}).
 Studies of the cosmic infrared background radiation (e.g., \citealt{hauserrr}) have shown that at least half of the total observable energy generated by stars has been absorbed by dust and reprocessed into infrared light. This result has indicated that the dust obscured star formation must have been much stronger at earlier epochs (e.g., \citealt{lagache05}; \citealt{franc08}).
%For example Takeuchi 05 has reported that at redshifts $0.5<z<1.2$ 70~\% of the star formation activity is obscured by dust.
Many authors report that at increasing redshifts, extreme star forming (SF) galaxies, which become more prominent, are heavily obscured by dust (e.g., \citealt{hop01}; \citealt{takeuchi05}; \citealt{ver07}; \citealt{sullivan01}).

Observations with infrared (IR) satellites like Infrared Space Observatory (ISO, \citealt{genzel00}) and the Infrared Astronomical Satellite (IRAS; \citealt{iras}), have revealed the existence of thousands of galaxies that emit strongly at IR wavelengths. With the advent of a more recent generation of IR satellites like the Spitzer Space Telescope (e.g., \citealt{spitzer}; \citealt{dole}; \citealt{frayer}) and AKARI (\citealt{matsu06}),  much larger and deeper samples of IR sources \citep{le05} with improved sensitivities were provided.

%\\
%Tu czemu 24um jest dfajne?\\
%\\

 As star formation is an extremely important factor in galaxy evolution, it is of crucial importance to study the properties of galaxies that are actively forming stars.
Of particular interest are mid-infrared (MIR) observations, as they trace the dust emission resulting from the heating by high-energy photons, and therefore they are invaluable in detecting sources of actively forming stars.
The MIR selection is affected by prominent spectral features. These features include strong emission from the polycyclic aromatic hydrocarbons (PAH) at rest-frame wavelengths of 3.3, 6.2, 7.7, 8.6, 11.3, 12.7, 16.3, and 17 $\mu $m, which can increase the number of sources within a flux-limited sample, or silicate absorption feature at 9.7~$\mu $m, which can cause a deficit in detection of the targeted sources.
 Obtaining knowledge about the way that different SF galaxy populations are placed in the context of the large scale structure (LSS) of the Universe can help to understand their origin and determine their future fate.
%Previous studies performed at these wavelengths ({\bf all based on observations from  Spitzer Space Telescope, e.g. \citealt{spitzer}, \citealt{dole}, \citealt{frayer}, albeit in different parts of the sky)} have revealed the existence of two distinctive populations of objects inhabiting different structures: the nearby dust-enshrouded star forming galaxies and very massive systems at higher redshift ($z \sim 2$) where the active phase has ceased at higher redshifts (e.g. \citealt{maglio}, \citealt{starikova}, \citealt{gilli}).

So far, in the field of clustering of mid-IR sources, many authors have relied on the data collected through Spitzer's 24~$\mu$m band, mainly because those wavelengths can provide a broad view of SF sources at different redshifts. The published results provided  great insight into the connection between the underlying dark matter distribution of IR galaxies and their evolution:
\begin{itemize}
%\item \citet{masci}
\item \citet{gilli} presented spatial clustering measurements of $\sim$800 galaxies from SWIRE MIPS 24 $\mu$m passband, observed at $z\sim1$. They found that depending on the different cuts of the total infrared luminosity ($L_{TIR}$) galaxies cluster in a different way: objects with higher $L_{TIR}$ cluster with greater strength. This implies that galaxies with higher star formation activity are hosted by more massive DMHs and denser environments than currently predicted by galaxy formation models.
\item \citet{maglio} used $\sim 1040$ SWIRE MIPS 24~$\mu$m selected sources brighter than 400~$\mu$Jy from two redshift intervals: at $z=[0.6-1.2]$ and $z>1.6$, to discover, that both populations are highly clustered. The high-z population was found to reside in very massive halos (comparable with local hosts of groups-to-clusters), while the low-z population resides on average in smaller DMH. 
Moreover, from the IR photometry it was revealed that both samples contain a similar mixture of star-forming galaxies and active galactic nuclei (AGN).
\item \citet{starikova} have analyzed properties of $\sim$ 20000 objects above a 310$\mu$Jy flux limit in MIPS 24~$\mu$m passband. They reported the existence of two populations of IR galaxies at $z_{med}\sim0.7$ and $z_{med}\sim1.7$. They reveal that both samples represent different populations of objects, which are found in differently sized DMHs, with the higher-$z$ objects residing in progressively more massive halos.
%\item {\bf \citet{dolley14} explored the clustering and halo masses of over $22000$ 24~$\mu$m selected star forming galaxies at 0.2$<z<$1.0. They repot weak clustering $$ find that galaxies with the highest star formation rates reside in } 
 %with corresponding correlation lengths ($r_{0}$, the clustering scale at which the correlation function falls below unity) equal to $\sim 5$ $h^{-1}$Mpc for the low redshift sample, and $\sim 8$ $h^{-1}$Mpc for teh high redshift one.
\end{itemize}
A general conclusion throughout the literature is that 24~$\mu$m sources are composed of two distinct populations of objects inhabiting different structures: the $z\sim1$ dust-enshrouded SF galaxies and very massive systems at $z \sim 2$ where the active phase has ceased at lower redshifts.
However, all of the above Spitzer studies differ amongst each other in many aspects, such as  the sample selection criteria (mostly based on the existence of counterparts in optical wavelengths), different methods of redshift estimation, and drastically varying surface density of the sources.
In light of those facts, it would be advisable to take an independent look at the 24~$\mu$m sources, from a perspective of another satellite, with a source selection method, which is not based on the existence of counterparts in optical wavelengths.
%TU CZEMU SPITZER SSIE?\\

The AKARI satellite was designed to carry
out IR observations with a sensitivity and resolution higher than that of preceding missions. It was launched by JAXA's
MV8 vehicle on February 22, 2006, carrying out, amongst others, a deep survey of the North Ecliptic Pole region (hereafter NEP), which we aim to use to explore the mid-infrared properties of galaxies, in particular, the evolution of clustering.
With a dense IR wavelength coverage, AKARI is well suited for delivering an independent look at the evolution of composition and clustering of different IR galaxy populations.
 In this research, we study the clustering properties of 24~$\mu$m galaxies from a perspective of an infrared satellite other than Spitzer, utilizing a new source selection method based on the usage of IR data alone.

 The paper is organized as follows: the description of the data, the method used to select the galaxies, and the way the redshifts were obtained can be found in Sect. 2; %Section 3 is dedicated to the description of the process leading to derivation of the photometric estimated of redshifts for the considered sources;
in Section 3 we present and summarize the methods used to calculate the angular correlation function and galaxy mass bias. Sect. 4 presents the results of application of those methods to the NEP data and we discuss and compare our findings with previous studies in IR and optical surveys. The summary and conclusions are given in Sect. 5.

Throughout this analysis we have assumed a flat $\Lambda$ CDM cosmology with $\Omega_{M}=0.27$, $\Omega_{\Lambda}=0.73$ and $H_{0}$100 h km $s^{-1}$ $Mpc^{-1}$.

\section {The data}

The NEP Deep sky survey covers an area of 0.4 sq.$\mbox{ deg}$ around 
the North Ecliptic Pole \citep{matsuhara}.
The data were obtained by the Infrared Camera (IRC; \citealt{onaka})
through nine near- and mid-infrared (NIR and MIR) filters, 
centred at 2~$\mu$m ($N2$), 3~$\mu$m ($N3$),
4~$\mu$m ($N4$), 7~$\mu$m ($S7$), 9~$\mu$m ($S9W$), 11~$\mu$m 
($S11$), 15~$\mu$m ($L15$), 18~$\mu$m ($L18W$), and 24~$\mu$m ($L24$), where W indicates that the bandwidths are wider than the others. 
%The long exposure times (from 1047 s for $N2$ filter to 261.8~s for $L24$ filter) resulted in reaching very deep into this region. 
%Table~\ref{tabbb1} summarizes the survey, where $\lambda _{\mbox{\tiny{ref}}}$ is the reference wavelength, $N_{\mbox{\tiny{sources}}}$ is the total number of detected sources in a specific bandpass ({\bf before any masking procedures}), $\mbox{mag}_{\rm lim}$ is the limiting magnitude of detected objects in a specific filter, and zero point stands for magnitude zero point used in brightness conversion procedures. 
The point spread function (PSF) has a beam size of $\sim 5$ arcsec (depending on the wavelength it varies between 4.4 and 5.8), which makes AKARI's imaging superior to that of other IR satellites.
The source extraction on FITS images was done using the SExtractor 
software \citep{sex}.
 A source is assumed to be detected if it has a minimum of 5 contiguous pixels above 
1.65 times the RMS fluctuations. Instead of allowing the program to 
estimate the background, weight maps were used (see \citealt{wada}). 
Photometry was carried out using SExtractor's MAGAUTO variable elliptical 
aperture with aperture parameters: Kron factor and minimum radius set to 
2.5 and 3.5, respectively. 
The zero magnitude points were derived from observations of standard 
stars \citep{tan} and are used to convert counts to magnitude by the 
photometry program.
The number of sources detected in individual filters differs significantly: 
far more sources are detected in NIR than in MIR.
The photometry resulted in detection limit of 19.3 $\mu$Jy at 24 $\mu$m ($L24$ filter). 
%obtaining a catalog depth of {\bf 1.4}~$\mu$Jy at 24 $\mu$m ($L24$ filter). 
The results of this procedure were downloaded from the official AKARI 
Researchers Web Page\footnote[1]{http://www.ir.isas.jaxa.jp/ASTRO-F/Observation/}, however, we have conducted security check runs of 
SExtractor, and the parameters obtained from this independent run were 
used in the subsequent analysis, after confirming that the basic results 
were consistent with the original catalog. 

\subsection{24 $\mu$m galaxy sample selection}
%In order to work on a purely 
 The first attempt to separate stars and galaxies within the NEP data was presented in \citep{ja}, where we exploited the Support Vector Machines (SVM) algorithm using all available color information for AKARI objects. Here, we employ the same method, this time aiming at selection of a pure 24 $\mu$m galaxy sample, which we use later on to explore the clustering properties of these kinds of objects. % To separate stars and galaxies within the NEP data we adopt a technique  
In this subsection,;l;l; we give a brief summary of the principles governing the selection method and then we describe how it was applied to obtain the best and most secure results.
\subsubsection{Method}
To separate stars and galaxies we employ the Support Vector Machines (SVM) algorithm, a supervised method based on kernel methods (\citealt{st}), allowing for pattern recognition within the provided data. The advantage that the SVM algorithm has over other algorithms is the ability to use all available information simultaneously, which has proven to be of great use in astronomy (e.g., \citealt{wozniak}; \citealt{zz}; or \citealt{hc}; \citealt{ja}; \citealt{my}). Here, we  outline the basic idea behind the method, however, for a more in-depth description see \citet{hsu} or \citet{crist}.
%In general, every object can be described by a vector containing its characteristic features. 
The task of the SVM algorithm is to divide the data points into two (or more) subsets according to the previously chosen adumbrative samples of desired classes by a supervisor, and to create a separating hyperplane, which  serves as the decision boundary for the data that we want to classify.
 To train the SVM algorithm means to input a feature vector for each object in training example, i.e., quantities that describe the properties of a given class' object, so we are mapping the input data from the input space $X$ onto a feature space (which can consist of an infinite number of dimensions) $H$ using a nonlinear function:
$ \phi\!:\!X\!\rightarrow\!H.$ In the parameter space $H$ the function that  determines the boundary can be written as 
\begin{equation}
 f(x)=\sum_{i=1}^{n}\alpha_{i}k(x,x')+b,
\end{equation}
 where $ k(x,x')$ is the kernel function returning an inner product of the mapped vectors, $\alpha_i$ is a linear coefficient, and $b$ is a perpendicular distance, which translates the boundary in a given direction.

%More specifically, SVM searches for a boundary, which will separate the training data points with the largest gap possible between each data point and this boundary. 
%In a non-linearly separable cases, we need to account for some amount of missclassifications: the classifier needs to allow for some degree of violations of the condition in order to create a boundary that will not confine too tightly to the training points. This way the classifier does not overfit the data, and, when applied to the sample of objects which class is yet unknown (the test sample), the boundary will be much more efficient in the segregation.

%We can use a mapping function, called a classifier, to transfer feature vectors into discriminant ones, which contain likelihoods of the given object to belong to one of the considered classes.
%The Kernel function, which are used to map input vectors non-linearly into a high dimensional parameter space and construct an optimal separating hyperplane.
%When applied to the NEP data directly
\subsubsection{Application to the NEP data}

In \citet{ja}, this method was directly applied to the NEP Deep sources with full photometric information.
%\citealt{ja} used this method directly to the NEP data set, which contained the full photometric information.
With the AKARI IRC flux measurements, we built a 6D parameter space  using the following color indices: $ N2-N3, N3-N4, N4-S7, S7-S11, S11-L15, L15-L18$. We excluded the {\it SW9}
and {\it L24} filters in order  to classify as many objects as possible,   since the amount of sources detected in those passbands was the smallest. Then, samples containing stars and galaxies, chosen by their stellarity parameter ($sgc$) value measured in NIR, were used to train the SVM and obtain its classifier.
The results reached the total classification accuracy of 93~\%, with specific accuracies of 98~\% for selecting stars and 90 \% for selecting galaxies. The results were tested with the auxiliary optical identifications and with the source count models, all of which proved that the classification based on infrared data alone is very efficient.
Therefore it was a natural next step to test it on the sources that were missing measurements due to the masking procedure and/or actual physical dropout properties.\\

To create a catalog of 24 $ \mu \mbox{m}$ AKARI galaxies. we performed a typical four-step routine for the application of SVMs to the classification task, as follows.\\
1. Manual selection of subsets of objects, which are representative to their predefined class, and   serve as a training basis for creating a classifier.\\
2. Each training source has to be described by its discriminating properties; in other words, for each training example there has to be a corresponding feature vector.\\
3. Selection of the kernel function. In this research we only use the radial basis function.\\
4. Training the algorithm to learn how to distinguish objects by creating a separation hyperplane described by two adjustable parameters ($C$, $\sigma$).\\

To complete these steps, firstly, as a training sample we  used previously classified stars and galaxies detected in all the AKARI passbands.
Then, we constructed the feature space  from seven IR colors.
% (six dimensional one from Solarz et al. 2012 was extended by adding the $L18W-L24$ values, which were missing for many objects from the full sample.
Based on the training sources we  created a classifier by performing a grid search of two free parameters ($\sigma$, responsible for the shape of the hyperplane, and $C$, governing the amount of missclassifications for which we allow)  and a tenfold cross-validation technique. The final pair of parameters ($(C,\sigma)=(10^{0},10^{-1}))$ was chosen based on the best accuracy of correctly classifying stars and galaxies within the training sample.
 The accuracy is defined as a ratio of all galaxies and stars whose nature was properly recognized by the classifier to the total number of considered objects. The final classifier was applied to all sources detected in the 24~$ \mu \mbox{m}$ band (2208 objects\footnote[3]{The number corresponds to the amount of objects left after masking unreliable regions of the image.}) to infer their respective classes.

% To this aim, to train the algorithm we have used the full-photometric stars and galaxies previously classified by the SVM, and we test it against the raw 24 $ \mu \mbox{m}$ based catalog of AKARI sources.
%we used the resultant samples of stars and galaxies, which were previously classified through the first application of the SVM to the data with full-photometry as training sets for the raw 24 $ \mu \mbox{m}$ based catalog of AKARI sources.}
 The resultant classifier exhibited a 83.89~\% total accuracy, with the true galaxy rate (TGR) and true star rate (TSR) for selecting stars and galaxies as 97.69~\% and 66.40~\%, respectively.
The specific accuracies are defined as follows:
$$TGR=\frac{TG}{TG+FS},$$
$$TSR=\frac{TS}{TS+FG},$$
where $TS$ stands for a true star (actual star identified as a star by SVM), $TG$ is a true galaxy (actual galaxy identified as a galaxy by SVM), $FG$ is a false galaxy (actual star classified as a galaxy by SVM) and $FS$ is a false star (actual galaxy classified as a star by SVM).
% The relatively low star selection rate during the training process is to be expected because of the nature of the classification process.
%Stars visible in the NIR get gradually dimmer to finally become undetectable when observed at longer wavelengths. Therefore in a 24 $\mu \mbox{m}$ band, we expect to detect a mixture of sources which contains very few real stars. 
As the classifier was built with a primary objective to select galaxies, any objects with rough spectral shape that was too different from a galaxy was classified as a so-called star. That however does not mean that such sources are actually stars.
The primary criteria for classification of an object were not only infrared colors, but also measurements of the sources' compactness (see \citealt{ja}).  That is why in addition to stars this sample could contain, for example, compact galaxies and AGNs. Therefore this group of objects shall be called a reject sample. As the class of those sources may be of a nature that we have not considered here, we left them out of the analysis to preserve the purity of the galaxy sample.
The rejected objects' number is 799, and, as stated before, this sample contains a mixture of stars and possible galaxies displaying unusual properties that have caused their displacement with respect to the separation hyperplane. 
%The classifier can be expanded to differentiate between more classes, however 
%, and contains all objects with unusual spectral features that differentiate them from galaxies, which are not a subject of this analysis. 

%in a parameter space constructed from a combination of infrared fluxes ranging from 2 to 24 $\mu$m, 

%and the corresponding statistical error of the TSR becomes large.
 %Among them we can expect to see very cold, old stars, which may be completely invisible in the NIR wavelengths. However, since the shape of the spectrum remains similar, the classifications which we can get can be considered accurate.

%The classifier, trained in the way described above, was applied to the full 24 $\mu \mbox{m}$ catalog of 
The resultant galaxy sample contains sources that are almost always detected also in {\it NIR-N} passbands. This suggests that our sample does not include a higher redshift fraction of intensively SF galaxies expected at $z\sim 1.6$ and $2.7$ (see, e.g., \citealt{houck}), a population that as a result of   high dust content is obscured in optical and near-infrared part of the spectrum. Nevertheless, as we report in the later sections, we find tentative evidence for a separate high redshift population of these kinds of  galaxies.
Estimation of the contamination of our samples is based on the number of objects belonging to the two classes that lie within the hyperplane boundary. If an SVM classified galaxy's (or star's) position in multicolor space has a separation boundary distance smaller than the error bar, it is treated as a possible misclassification. On this basis, the contamination in our galaxy sample is estimated to be 4.97 \%. After a rejection of the boundary region, the final catalog contains 1339 galaxies.
%, and those objects were rejectedexcluded from the resultant catalog.
%The mask 

\begin{figure}[!h]
\centering
\includegraphics[width=0.4\textwidth]{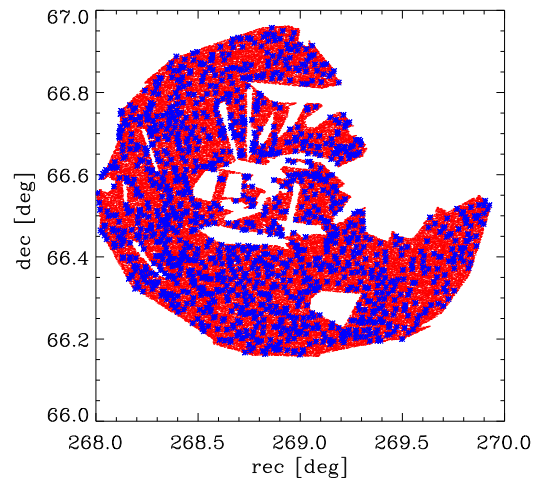} 
\caption{Position map of the MIR galaxy sample chosen by  means of the Support Vector Machines algorithm (blue points). Red points mark the shape of the mask of the field}
\label{positions}
\end{figure}

Figure \ref{positions} shows positions of galaxies in the final 24-$\mu$m selected catalog within the masked NEP region.
The masking procedure included the removal of bad columns, high-noise areas and overexposed bright objects (like the Cat's Eye Nebula, which covered almost one-third of the field in 24 $\mu $m image.)

We estimate the completeness of the derived catalog empirically to be 150~$\mu$Jy, by defining the limit as a point of the maximum of the differential number counts of sources.
%On the $ {\rm \log N-\log S}$ scale the differential counts do not show a sharp peak, but instead they exhibit a plateau extending from $\sim$ 150 $\mu \mbox{Jy}$ to $\sim$ 350 $\mu \mbox{Jy}$, which could mean that we are dealing with a flux uncertainty of $\sim$ 200 $\mu \mbox{Jy}$.
%That is why we choose out completeness limit to the certain $150 \mu m$ cut.
%%%%%%%%%%%%%%%%%%%%%%%Tu walnac moze wykres NC?

\subsection{Redshift estimation}

%96.11\% of sources used in our analysis have the measurements obtained only in the infrared, and do not possess any counterpart in public uncatalogued.
%In order to relate the angular and spatial correlation functions we need to know the redshift distribution of the considered sources. 
%******************************
Since follow-up spectroscopic measurements are time consuming and, for the AKARI NEP Deep Field, the existing measurements are very sparse (which is addressed further in the text), and we have to rely on the photometric estimation of the redshifts.
%technologically and time expensive
 The usual photometric redshift estimation codes are based on fitting the abundance of optical templates for sources, like the LePhare routine \citep{lephare}. 
However, since the only measurements available for our sources are performed in the $NIR$ and $MIR$, we had to approach this task in an alternative way.
To this aim, we ran the \emph{Code Investigating GALaxy Emission} (CIGALE;  \citealt{noll09}), a spectral energy distribution (SED) fitting routine. 
CIGALE was not developed as a tool for estimation of photo-$z$ but since it uses a large number of models covering the whole spectrum including IR, it may be expected to provide $z_{\mbox{\tiny{photo}}}$  using only the infrared part of spectra \citep{malek12} with satisfactory quality.
The CIGALE technique uses models describing the emission from a galaxy in the wavelength range from far-UV to far-IR. 
Models of emission from stars are given either by \citet{maraston05} or \citet{fioc97}.
The absorption and scattering of star light by dust, the so-called attenuation curves for galaxies, are given by \citet{calzetti00}.
Dust emission is characterized by a model proposed by \citet{dale02}. 
In this model, the IR part of an SED is given by a power-law distribution
\begin{equation}
 {\rm d}M_{{\rm d}}(U) \propto U^{-\alpha_{{\rm SED}}}dU,
\end{equation}
where $M_{\rm d}(U)$ is the mass of dust heated by a radiation field $U$, and $\alpha_{\rm SED}$ is a heating intensity.

 To
find a photometric redshift with CIGALE, it is necessary to shift a galaxy spectrum to many different redshifts and then to run CIGALE to fit SEDs for the same galaxy with a number of combinations of different redshift values. We decided to estimate photometric redshift for the AKARI NEP sample in the redshift range from $ z_{\rm{min}}=0.01$ to $z_{\rm{max}}=3.00$ with a step $\delta z$=0.02.
The final redshifts assigned by CIGALE are based on the minimal $\chi^2$ value of the fit with 7 degrees of freedom (for details, see \citealt{noll09}). 
%{\bf The obtained redshift distribution is presented in Fig\sim\ref{dz} as a solid thick line.}
In the same sample of 1339 galaxies,   it was not possible to  estimate photometric redshift ($\chi^2$ equal to  $99.99$) for only 4.85 \% of the sources. 
For 38.31~\% of the galaxies, the $\chi^2$ was lower than 10. 

\begin{figure}[!h]
\centering
\includegraphics[width=0.4\textwidth]{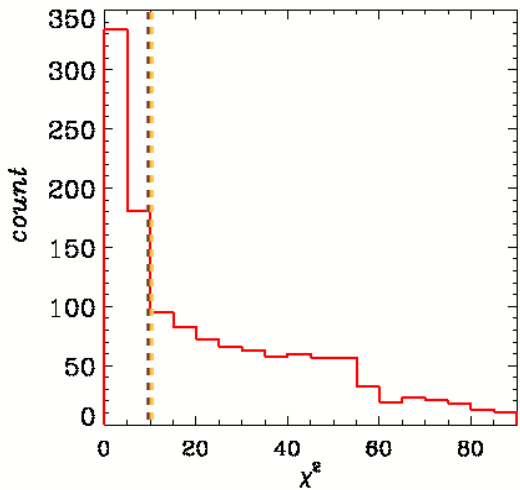} 
\caption{The distribution of the $\chi^2$ value obtained by redshift estimation analysis (shown in bins of 5.0). The vertical dashed line marks a cut at $\chi^2=10$, corresponding to the best SED fits used later for the analysis. Sources with $\chi^2$ values equal to 99.99 are not shown. }
\label{chi2}
\end{figure}

The distribution of the $\chi^2$ is shown in Figure~\ref{chi2}. The resultant redshift distribution obtained using CIGALE is presented in the Fig.~3, where it is compared to the other estimations previously published in the literature.
The detailed description of the application of CIGALE to AKARI data is discussed at length in \citet{kaska14}.
%(a) (solid line) and compared with the ones found in the literature (\citealt{starikova} (dotted line), \citealt{maglio} (short-dash line), \citealt{gilli} (long-dash line), \citealt{desai} (with (dash-dot line) and without (dash-triple dot line) spectroscopic redshifts) and the one obtained by \citep{goto10} for a NEP data based on the optical follow-up observations (Fig.\sim\ref{nofz}(b)).
%The two distributions show similar features, with the peak of the distribution 

%*******************************

%%%%%%%%%%%%%%%%%%
%Tu tabelka z podsumowaniem katalogu?
%{\bf The redshift distribution of the AKARI NEP Deep galaxies obtained by the CIGALE code shows}
 Previous studies of photometric redshift estimates of the 24 $ \mu \mbox{m}$ selected sources have identified two peaks in the redshift distribution.
Depending on the flux density limits, many studies have reported a rise in the detection of sources at $z < 1$  peaking between $z \sim 0.3 $ (e.g., \citealt{desai}) and $z\sim 0.8$ (\citealt{perez-gonzalez05}; \citealt{le05}; \citealt{caputi}).
%on the other hand
%, \citep{desai} determines the position of the primary peak to be located at $z \sim 0.3 $. 
 At these redshifts there are no prominent spectral features in the 24 $ \mu \mbox{m}$ band except for the $\sim16$ and 17 ${\rm  \mu m}$ PAH emission bands. %, therefore the increase in detections is possibly caused by either evolution in the luminosity function or by the fact that the probed volume increases with redshift.
Some authors (e.g., \citealt{caputi}) have reported that between the redshifts of $z\sim 0.8$ to $z \sim 1.6$ the number of detected 24 $ \mu $m objects decreases significantly, possibly because of the deep silicate absorption line features at 9.7 $ \mu \mbox{m} $ rest frame causing some sources to fall out of the flux detection limit.
However, \citet{desai} report a tentative detection of a distribution peak at $z\sim 0.9$, which has not been detected in  previous studies. This distribution peak can be attributed to the 12.7 $ \mu \mbox{m} $ PAH feature and 12.8 $ \mu \mbox{m}$ [Ne$\textsc{i}\textsc{i}$] emission line, which at this redshift is seen in the 24 $ \mu \mbox{m}$ passband. 

An additional peak predicted by models (e.g., \citealt{lagache}), which occurs at $z\sim 2$, can be attributed to sources dominated either by an AGN, or as a consequence of PAH emission features at $\sim$ 8$\mu \mbox{m}$ (e.g., \citealt{caputi}). 

%At redshift $z\sim 2$ many authors report an existence of the secondary peak, whoever due to the lack of the spectroscopic measurements cannot be explicitly explained; however a strong PAH emission feature at $8 \mu m$ could be responsible for galaxies to be more prominent in $24 \mu \mbox{m}$ at this redshift.

%{\bf The two populations of 24 $ \mu \mbox{m}$ sources existing at redshifts $\sim 1$ and $\sim 2$ have been reported not to be each other's high/low redshift counterparts \citep{maglio}}. and probably should be treated as different source populations.}
% The primary peak is located at redshifts lower than 1, reporting the position to be at, e.g. $z\sim 0.8$ \citealt{perez-gonzalez05}, $z\sim0.7$ \citealt{le05}, where the difference in determining the precise location is attributed to the cosmic variance; moreover, Desai 2008 reports the position of the primary peak to be positioned at $z \sim 0.3 $; the difference  and is created by the strong luminosity and/or density evolution occuring within the IR-luminous galaxies on those redshifts.
%{\bf Then, armed with this knowledge we estimate the photometric redshift distribution of the AKARI NEP Deep 24 $\mu$m galaxies}
The results of our photometric redshift derivation using the CIGALE code reveals results that are in agreement with these studies (see Fig. 3). We observed two expected peaks in the photo-$z$ distribution, one located at $z\sim 0.6$ and the weaker secondary peak located at $z\sim1.2$.
Moreover, we detect a noticeable rise in the counts at $z\sim 2.4$.
%While this peak cannot be confirmed by \citet{goto10} {\bf (see Fig. 3b)} due to the fact that their redshift estimation did not extend to objects with redshifts higher than $z\sim 2$, in the studies of \citet{desai}, the existence of a peak at this redshift is evident.
\begin{figure}[ht]
    \centering
    %\subfigure[a]
    {
        \includegraphics[width=0.4\textwidth]{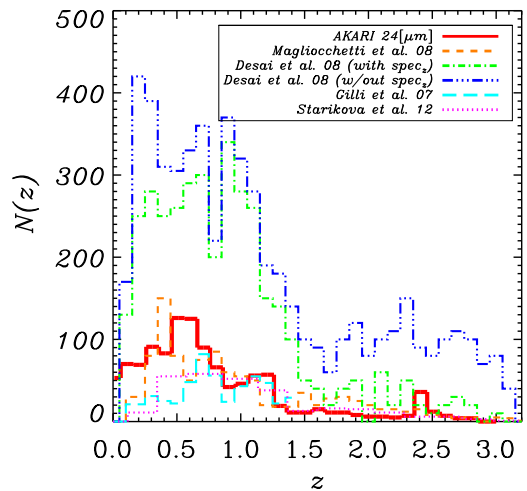}
%        \label{111}
    }
    %\subfigure[b]
    %{
     %   \includegraphics[width=0.4\textwidth]{histo_goto.eps}
%        \label{222}
    %}
    \caption{Redshift distribution of the AKARI NEP Deep Field MIR selected galaxies derived by CIGALE code (solid line) together with those published in the literature (dashed: \citealt{maglio}; dash-dot-dashed and dash-triple-dotted: \citet{desai}, for objects with and without spectroscopic redshifts; large-dashed: \citealt{gilli}; and dotted: \citealt{starikova}). }
%Right panel: redshift distribution of the  AKARI NEP Deep Field MIR selected galaxies derived by CIGALE code (solid line), compared with the one derived by \citet{goto10} indicated by the dashed line. }
    \label{dz1}
\end{figure}

%\subsubsection{Spectroscopic follow-up observations}

\begin{figure}[!h]
\centering
\includegraphics[width=0.4\textwidth]{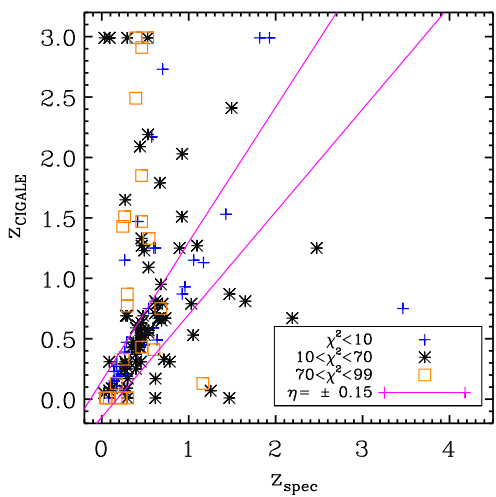} 
\caption{ Comparison between $z_{\rm{CIGALE}}$ and $z_{\rm{spectro}}$. Different symbols correspond to different $\chi^2$ values: blue crosses denote objects with $\chi^2 < 10$, black asterisks  objects with $10 \le \chi^2<70$,  and orange squares objects with $70 \le \chi^2<99$. Solid lines are for $z_{\rm{CIGALE}}=z_{\rm{spectro}} \pm 0.15(1+z_{\rm{spectro}})$. }
\label{ce}
\end{figure}

\citet{murata13} recently released a catalog with measurements of spectroscopic redshifts for a sample of 307 objects within the AKARI NEP Deep field obtained by DEep Imaging Multi-Object Spectrograph (DEIMOS) in Keck II telescope (\citealt{deimos}; Goto et al. in prep).
Since the best way to estimate the accuracy of the photo-$z$ measurement procedure is to check it with directly measured spectroscopic redshifts, we have cross-correlated our catalog with that of \citet{murata13}. 
We have found 149 counterparts from the spectroscopic catalog. Six of them resulted in CIGALE finding no believable redshift ($\chi^2 > 99$). %{\bf, and in those cases we substitute the no-redshift with the spectroscopic one}.
To check the reliability of the photometric estimations we estimate the amount of the catastrophic errors (CE) for sources, following \citet{smetnydred}. 
A CE occurs when the value of
\begin{equation}
{\rm \eta=\frac{|z_{spec}-{\rm (photo-}z)|}{1+z_{spec}}}
\end{equation}
exceeds $0.15$.
We estimate this value for several limiting $\chi^2$ cuts  to determine the value that is the most reliable for a clustering analysis. Figure~\ref{ce} shows the comparison between the derived photometric redshifts and spectral measurements, whereas Table~\ref{ce1} presents the results of this procedure.
The amount of the CE reaches $18.37\%$ for  $\chi^2 < 10$ and then increases significantly with the increase of the $\chi^2$ value until it reaches a value of almost twice as high for the $\chi^2 < 70$. \\
\begin{table}[!h]
\caption{Comparison of ${\rm z_{CIGALE}}$ vs ${\rm z_{spec}}$ as a function of $\chi^2$. $\eta$  - percent of CE in the sample.}
\begin{center}
\begin{tabular}{l l l l }
\hline\hline
$\chi^2$ & ${\rm N_{gal}}$ &  ${\rm N_{CE}}$ & $\eta$ [\%] \\ \hline
70 & 126 & 44 & 34.92 \\
50 & 112 & 39 & 34.82 \\
30 & 73 & 21& 28.77 \\
20 & 65 & 18 & 27.69\\
10 & 49 & 9 & 18.37 \\
\hline
 \end{tabular}
 \end{center}
 \label{ce1}
\end{table} 

%{\bf OBRAZEK Z FITAMI SPECTROW DLA ROZNYCH CHI2 Z UZASADNIENIEM, ZE W ZASADZIE TE Z 70 TO SA W TRABKE}
\begin{figure*}[ht]
    \centering
    %\subfigure[a]
    {
       \includegraphics[width=0.45\textwidth]{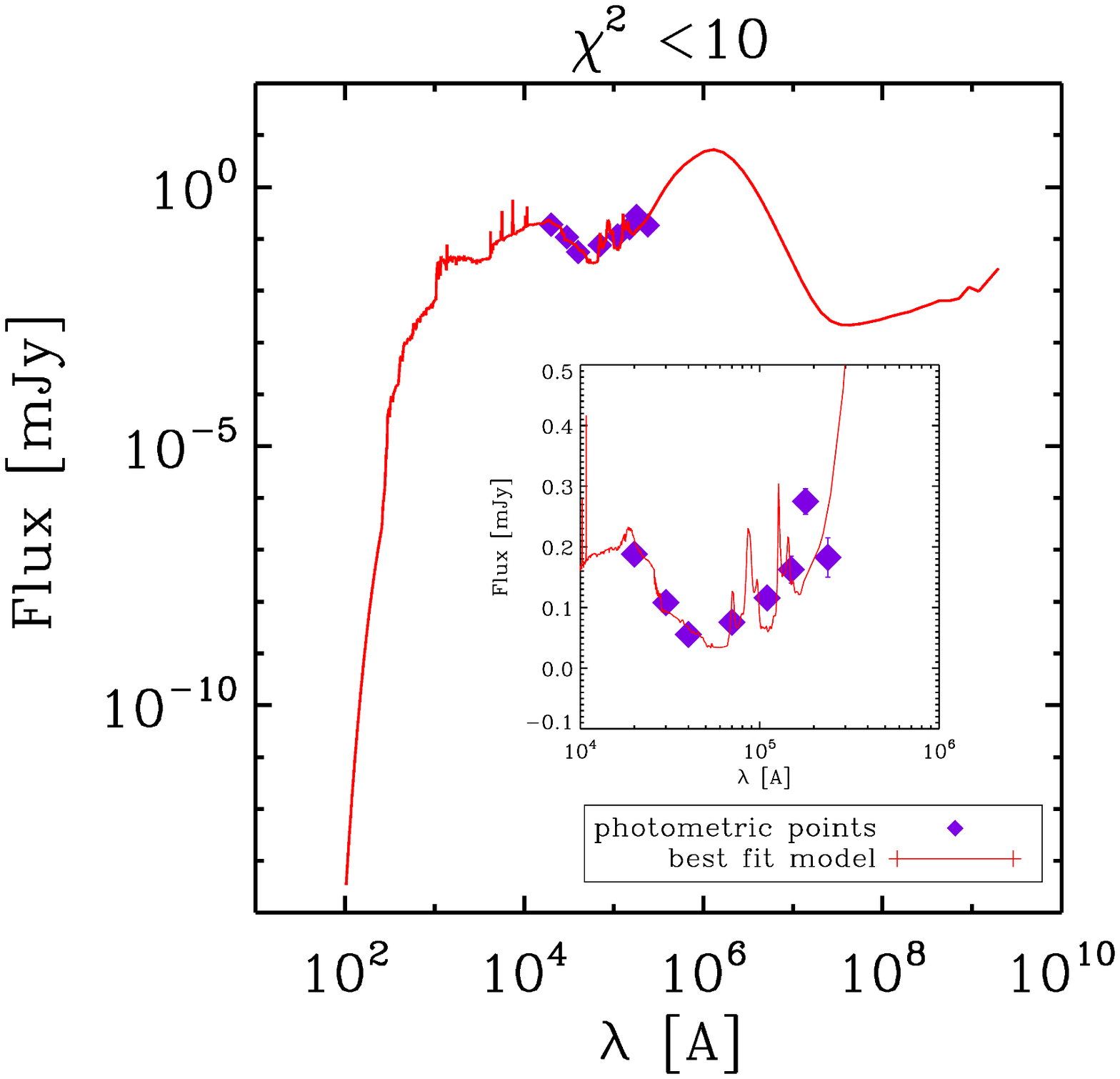} 
%        \label{111}
    }
    %\subfigure[b]
    {
        \includegraphics[width=0.45\textwidth]{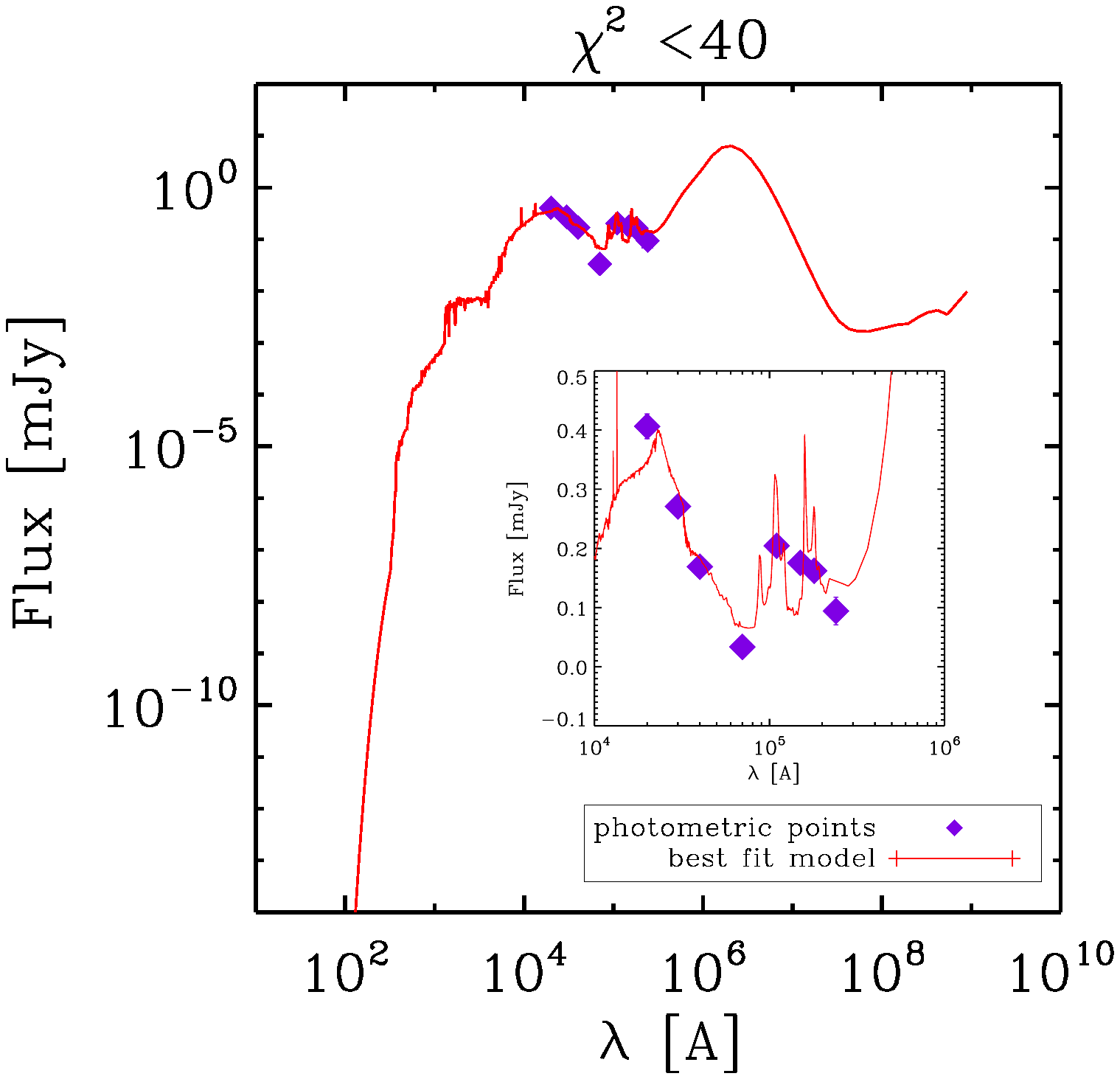}
%        \label{222}
    }
 {
        \includegraphics[width=0.45\textwidth]{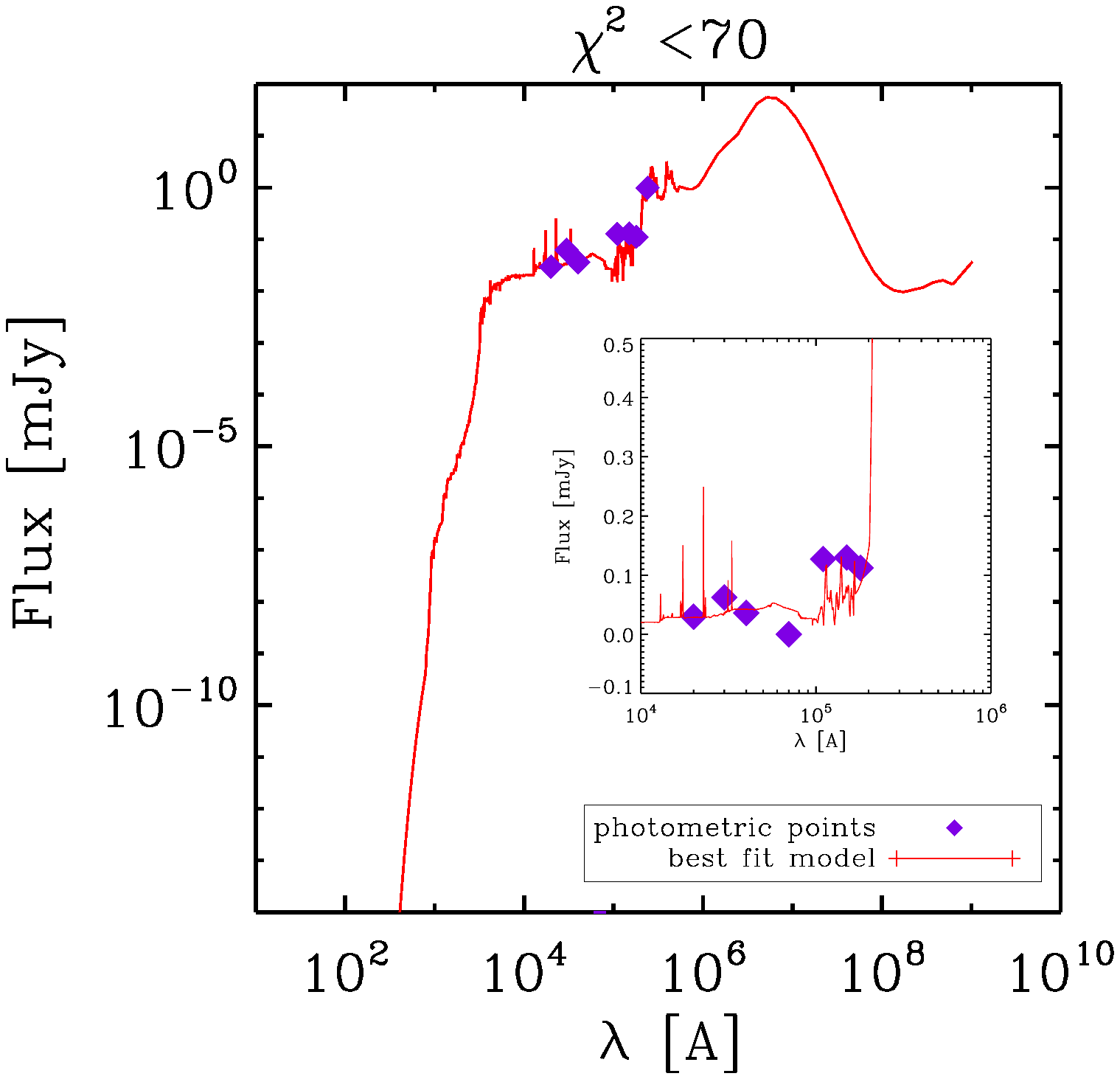}
%        \label{222}
        \includegraphics[width=0.45\textwidth]{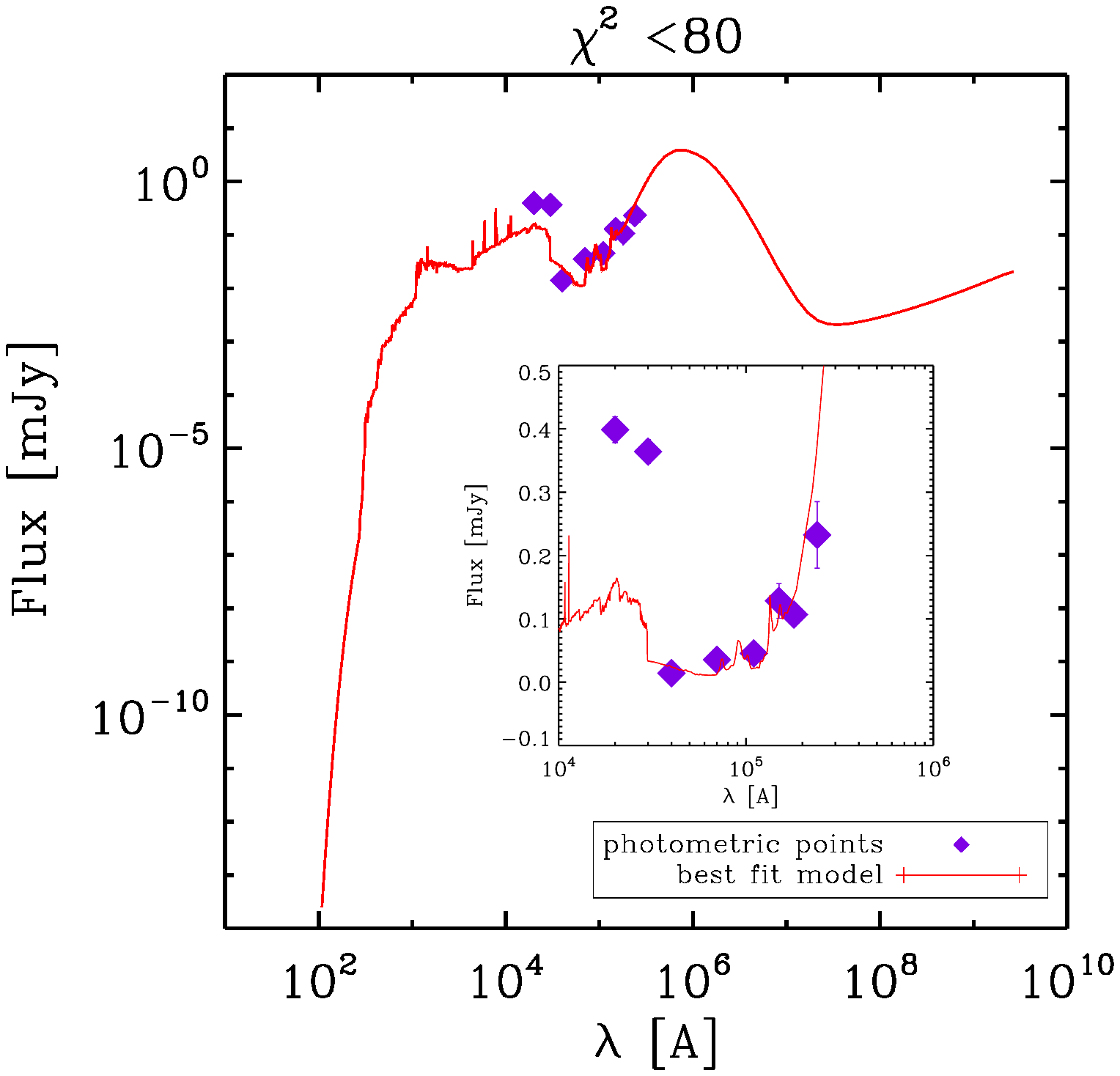}    

}
    \caption{Examples of the CIGALE fitted SED's to the AKARI NEP Deep data for different $\chi^2$ values. Each plot has a close-up on the photometric data in the linear scale. }
    \label{sed}
\end{figure*}

\begin{figure}[!h]
\centering
\includegraphics[width=0.4\textwidth]{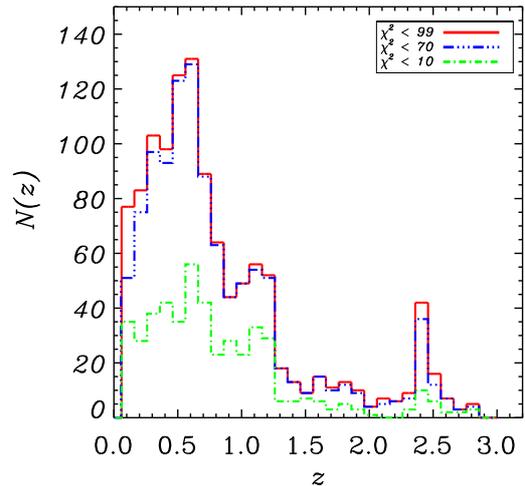} 
\caption{The redshift distribution of the AKARI NEP Deep sources obtained by the CIGALE code as a function of the limiting $\chi^2$ value. Distribution of all sources, with a successfully calculated $z_{\rm photo}$ is represented by the solid line, distribution of sources with photometric redshift measurements on the level of $\chi^2<10$ by dash-dotted line, and $\chi^2<70$ by dash-triple-dotted line.}
\label{hc}
\end{figure}

%{\bf Fig.\sim\ref{ce}  }
In Fig. \ref{sed} we present the examples of the resultant SEDs obtained with different $\chi^2$ values. 
It is clear that for values from $\chi^2 \sim 40$ up to the $70$ we are still dealing with the very good fits to the data.
%, mostly with one outlying point, and the difference between those values depends only on the distance of the data point to the predicted value.
 The fits with values of $\chi^2 \sim 80$ no longer seem reliable. 
The redshift distributions of the sources included within a specific $\chi^2$ cut are presented in Fig. \ref{hc}.
%\begin{table}[ht]
%\caption{Summary of the resultant uncatalogued}
%\begin{center}
%\begin{tabular}{ccccc}
%\hline\hline
%$\chi^2$  & ${\rm N_{\rm gal}}$& A [$\times {\rm 10^{-3}}$] (at ${\rm 1^\circ}$)& $\gamma$  & ${\rm r_0 [h^{-1}\mbox{Mpc}]}$\\ \hline 
%99&1274&$2.25\pm0.01$&$1.77\pm 0.05$&$4.93\pm0.49$\\%bylo 1292 N_gal
%10&611&$4.46\pm0.03$&$1.79\pm0.04$&$7.33\pm0.56$\\
%70&1220&$1.83\pm 0.06$&$1.75\pm 0.07$&$4.54\pm 0.72$\\
%\hline
%\end{tabular}
%\end{center}
%\label{podsum}
%\end{table}
%TUTAJ O SPEC Z??
 For the most reliable results, we decided to replace the photometric redshifts for objects that have a spectroscopic measurement from the Murata et al. catalog, and assigned them the value of $\chi^2=0$. This procedure increased the number of sources from 511 to 611 for the ${\rm \chi^2<10}$ sample, from 1195 to 1220 for the ${\rm \chi^2<70,}$ and 
from 1268 to 1274 for the full catalog without the galaxies for which the CIGALE code was unable to retrieve the redshift.
% We keep the purely photometric measurements as a control for the following correlation function calculation (see Appendix, Sec. 6).

\section{The angular correlation function}
 Galaxy clustering is traditionally analyzed in the first place by the means of the two-point correlation function, $\xi(r)$, which is treated as an above random excess probability of finding a galaxy within a certain distance $r$ from another galaxy. Often it is approximated by a power law $\xi(r)=(r/r_0)^{(-1/\gamma)}$.%, where $\gamma$ is the slope of the function, and $r_0$ is the correlation length. }
\subsection{Estimator}
To compute the angular correlation function, we use the following estimator introduced by \citet{ls}:
\begin{equation}
{\rm \omega(\theta)=\frac{DD(\theta)-2DR(\theta)+RR(\theta)}{RR(\theta)}}
\end{equation}
where ${\rm DD(\theta)}$ is the number of galaxy-galaxy pairs, ${\rm RR(\theta)}$ is the number of random-random pairs, and ${\rm DR(\theta)}$ is the number of galaxy-random pairs within an angular bin of separation on the sky ${\rm \theta}$.

We use $10^5$ homogeneously generated random points within the FOV and overlay the photometric mask with the identical features as the real FOV.
Because of the masking of  bad sectors of the FITS file for the 24 $\mu \mbox{m}$ passband, the effective number of galaxies used for the calculation was significantly reduced (by $\sim 36$ \% of the original number of sources). 
% with the exception of the areas cut out during the masking procedure.%, to minimize both the statistical errors and uncertainties introduced by the inconsistent(?) geometry of the field.
 The correlation function was computed in the angular bins of width $\delta \log \theta=0.1$. 
\subsection{Errors}
The uncertainties arising from computation of the correlation function through the Landy-Szalay estimator are
\begin{equation}
{\rm \delta \omega(\theta)=\frac{1+\omega(\theta)}{\sqrt{DD(\theta)}},}
\end{equation}
which accounts for the standard deviation in the pair counting along with the intrinsic variance. However it does not include the error arising from the covariance of the correlation function at different
separations. There are a number of ways to avoid this problem; for instance,  using bootstrap resampling of the data (e.g., \citealt{boot} ), mock catalogs (e.g., \citealt{pollo05}), jackknife resampling (e.g., \citealt{gaztanaga}; \citealt{ross}; \citealt{scrantan}), and many others.   We employ the jackknife resampling of the data to calculate  errors. To this aim we divided the observed field into ten equally sized areas and computed the correlation function ten times, each time leaving one area out. Then, the errors were estimated from Eq.~\ref{sigmaerr},
\begin{equation}
{\rm \sigma^2=\sum_{i=1}^{10} [\omega_i(\theta)-\omega(\theta)]^2,}
\label{sigmaerr}
\end{equation}
where $i$ denotes a particular subsample iteration.
We calculated all presented uncertainties from here on  in this manner.

\subsection{Fitting procedure and integral constraint}

Usually a correlation function follows the power law
\begin{equation}
{\rm \omega(\theta)=A_{\omega}(\theta^{\bf{1-\gamma}}),}
\end{equation}
(e.g., \citealt{pee}). Every estimate of ${\rm \omega({\theta})}$ is penalized by the finite size of the surveyed area and referred to as the "integral constraint" (IC, \citealt{peegroth}). This effect causes a negative offset of the observed ${\rm \omega(\theta)}$ with respect to the true correlation function, ${\rm \omega_{T}(\theta)}$, by a constant
$${\rm IC=\int \int \omega_{T}(\theta) d\Omega_1 d\Omega_2 ,}$$
where ${\rm d\Omega_1}$ and ${\rm d\Omega_2}$ are the transverse solid angles of each pair.
 It can be approximated using the following formula:
\begin{equation}
IC=\frac{\Sigma RR(\theta)\omega(\theta)}{\Sigma RR(\theta)},\end{equation}introduced by \citet{roche}. After applying this correction, the correlation function can be written as ${\rm \omega(\theta)=A_{\omega}(\theta^{\bf{1-\gamma}}-IC)}$.
To determine the best-fit parameters ($\gamma$, $A$), we performed a ${\rm \chi^2}$ minimization including the full covariance matrix $C$, i.e.,
$${\rm \chi^2=[\omega(\theta)-(A_{\omega}\theta^{{\bf1-\gamma}}-IC)]^{T} C^{-1} [\omega(\theta)-(A_{\omega}\theta^{\bf{1-\gamma}} -IC)].}$$

\subsection{Limber inversion and space clustering}
Using the measurements of the angular clustering, we can infer the three-dimensional clustering properties based on the known redshift distribution via Limber's equation (\citealt{pee80}, \citealt{limb}). When dealing with small scales both angular and spatial correlation functions are well described by power laws (e.g., \citealt{dp}) and therefore Limber's equation can be presented in the form (\citealt{efstat}),
\begin{equation}
{\rm A_{\omega}= C_{\gamma} r_0^{\gamma}\frac{\int d_{A}^{1-\gamma} x^{-1}(z)(dN/dz)^2 dz}  {[ \int (dN/dz)]^{2}}},
\end{equation}
where 
$d_{A}$ is the angular diameter distance, $x(z)$ is the derivative of proper distance with redshift,
%\begin{equation}
%{\rm x(z)=\frac{c}{H_0}\int\frac{dz}{E(z)}}
%\end{equation}
%is the cosmology dependent comoving radial distance,
%\begin{equation}
%{\rm E(z)=\sqrt{\Omega_{M}(1+z)^3+\Omega_k(1+z)^2+\Omega_{\Lambda}},}
%\end{equation}
\begin{equation}
C_{\gamma}=\frac{\sqrt{\pi}\Gamma(\frac{\gamma-1}{2})}{\Gamma(\gamma/2),}
\end{equation}
 and dN/dz is the redshift selection function, which in case of our study is derived empirically from SED fitting (see Fig.~\ref{dz1}).
\subsection{ Galaxy bias} 
In order to relate galaxy clustering to dark matter clustering, \citet{kaiser} and \citet{bard86}  introduced a bias parameter, a quantity  describing the differences between the clustering of  a galaxy field and the underlying mass distribution, i.e.,
\begin{equation}
b^2(r,z,M)=\xi_{g}(r,z,M)/\xi_{m}(r,z),
\label{biaspar}
\end{equation}
where $\xi_{g}(r,z,M)$ is the correlation function of the investigated galaxy population, and $\xi_{m}(r,z)$ is the correlation function of the dark matter. Bias is dependent on scale ($r$), redshift ($z$) and objects' mass ($M$).
%In the regime where a linear relation between the fluctuations of mass traced by galaxies (${\rm\sigma_g}$) and mass ({\rm$\sigma_m$}) is assumed, bias parameter should fulfill: 
In terms of the assumed fluctuations of mass traced by galaxies (${\rm\sigma_g}$) and mass ({\rm$\sigma_m$}), the bias parameter should satisfy the  relation
\begin{equation}
\rm{\sigma_{(g,R)}=b\sigma_{(m,R)}}. 
\end{equation}
It can be presumed that $b$ is invariant as long as the scales are large enough. 
In this case it is customary to compute $\sigma_{R}$ for a representative scale $R=$8~${\rm  h^{-1}}$Mpc \citep{quadri}.
%Present day ($z=0$) mass fluctuations on those scales reach the values of 0.76 (Spergel et al. 2007)
To obtain the value of ${\rm\sigma_{(m,8)}}$ for different cosmic epochs, one can use a following formula:
\begin{equation}
\rm{\sigma_{(m,8)}(z)=\sigma_{(m,8)}(0)D(z),}
\end{equation}
where {\rm$\sigma_{(m,8)}(0)$} is a present day ({\rm$z=0$}) mass fluctuation, which on the scale $R$ of 8 {\rm$h^{-1}$}Mpc has a value of 0.83 \citep{planck1}; 
and
\begin{equation}
\rm{D(z)=\frac{g(z)}{g(0)(1+z)},}
\label{dz}
\end{equation}
where {\rm$g(z)$} is a normalized growth factor that describes the growth rate of the matter density fluctuations, and can be calculated, according to \citet{carroll92}, as follows:
%$$$$
%\scriptsize{
\begin{equation}
\begin{aligned}
%\begin{split}
g(z) &\sim \frac{5}{2} \Omega_{m}(z)[\Omega_{m}^{4/7}(z)-\Omega_{\Lambda}(z) +\\
&+\left(1+\frac{\Omega_{m}(z)}{2}\right)\left(1+\frac{\Omega_{\Lambda}(z)}{70}\right)]^{-1}.
\label{growthfact}
%\end{split}
\end{aligned}
\end{equation}
The value of the fluctuations of the galaxy density ($\sigma_{g,8}$) with respect to the assumed scale can be obtained by implementation of the correlation function slope $\gamma$ and correlation length {\rm$r_{0}$}, i.e.,

\begin{equation}
{\rm\sigma_{g,8}=\sqrt{J_2\left( \frac{r_0}{8 h^{-1} Mpc} \right)^{\gamma} },}
\end{equation}
where {\rm$J_{2}=72/(2^{\gamma}(3-\gamma)(4-\gamma)(6-\gamma)).$}

\section{Results}

Following the procedures described in the previous subsections we compute ${\rm \omega({\theta})}$ for the 24 ${\rm  \mu \mbox{m}}$ selected galaxy catalog. The angular distances were measured in bins of ${\rm \Delta \mbox{log} =0.2}$. We then fitted the data using a $\chi^2$ technique with $2\sigma$ clipping and determined the errors from the covariance matrix. 

%The results obtained for the full catalog (with the exception of the galaxies for which CIGALE was not able to compute the redshift, ${\rm \chi^2 > 99}$) together with the {\bf one obtained for the best fit sub-catalog} are presented in Fig. \ref{full}.
The results of the power-law fitting to the correlation function between the full catalog (with ${\rm \chi^2 > 99}$) and subcatalog of the best redshift quality (${\rm \chi^2 <10}$) show, that the two samples differ significantly. 
The full catalog showed $r_{0}=4.93\pm0.49$ $h^{-1}\mbox{Mpc,}$ while the subcatalog showed $r_{0}=6.47\pm0.56$ $h^{-1}\mbox{Mpc}$.

%the exception of the galaxies for which CIGALE was not able to compute the redshift, 
% together with parameters obtained for the sub-catalog of the best spectral quality { (${\rm \chi^2 <10}$), \bf . 
%are summarized in Tab.~\ref{chionly}.
 The low $r_0$ value for the sample containing all $\chi^2$ values but the failing ones could be a result of the fact that it contains galaxies with lower fluxes, which would dilute the clustering signal.

%{\bf zrobic appendix}
% Ale pozniej o niej nie wspominam. zagadka! moze do appendixu?
%In what follows, we perform an analysis of the subsamples of sources with the most secure redshift measurements (we reject the sources with the values of $\chi^2$ of the photometric estimation higher then 10) and with the $\chi^2 < 70$.
%If we take into the consideration the samples with the limiting $\chi^2$ cuts of 70 and the basic sample with the only objects removed being the ones with $\chi^2\sim 99$, we see very little change in the best fit parameters. Since there is a very small difference in the number of objects within those samples, this is something we would expect. 
The difference is significant between the fit values for the subsamples with $\chi^2$ smaller then 10 and the full catalog.
 Depending on the $\chi^2$ cut we see drastic changes in both the strength and in the slope of the clustering. 

\begin{figure}[!h]
\centering
\includegraphics[width=0.4\textwidth]{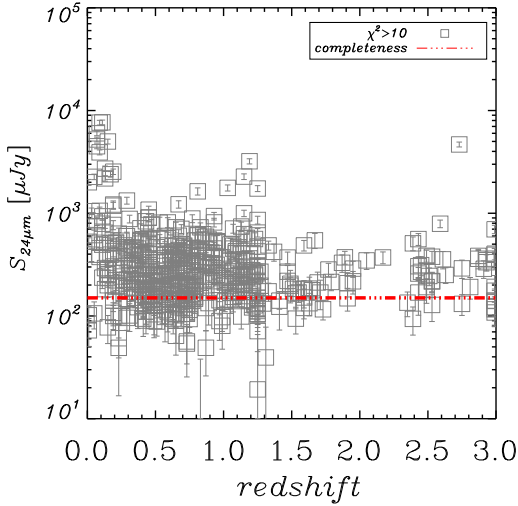} 
\caption{ 24~$\mu$m flux of objects from $\chi^{2}<10$ sample as a function of their redshift. The flux limit, marked as a red line, was derived from maximum of differential number counts of detected sources.}
\label{f24z}
\end{figure}

The most secure galaxies (with redshift estimation on the level of $\chi^2<10$) exhibit a much larger correlation length when compared with the whole sample. The $\chi^2<10$  galaxies are concentrated around $z\sim0.7$ (dash-dotted line on Fig.~\ref{hc}). 
The brightness distribution of those sources reveals that this sample consists mostly of the objects with ${\rm S_{24 \mu m}} > 500$~$\mu$Jy. 
This result could indicate that galaxies with the most secure redshifts are the most luminous objects within the AKARI sample.
In the following analysis, we only focus  on this subsample of galaxies.
 We present the 24~$\mu$m flux as a function of redshift for the $\chi^{2}<10$ sample in Fig.~\ref{f24z}.
\subsection{High- and low-redshift population sample}
%The redshift distribution based on the photometric estimation revealed several features.
 The galaxies with a successfully calculated $ z_{\rm CIGALE}$ on the goodness of fit level of $\chi^2<10$ were divided into four redshift ranges based on the shape of resultant distribution: 1) the low-$z$ sample composed of sources with $z\leq 0.5$; 2) the low-intermediate sample with $0.5<z\leq 0.9$, which includes the primary and most evident peak of the distribution which, as stated before, could be a result of 16.3 ${\rm  \mu m}$ and/or 17 ${\rm  \mu m}$ PAH emission features passing through 24 $\mu$m passband; % where, as stated before, we observe either an evolution in the luminosity function and/or at those distances we probe a larger volume resulting in the increase of the detected sources;
 3) the intermediate sample with $0.9<z\leq 1.3,$ including the secondary peak attributed to the 12.7~${\rm  \mu m}$ PAH and/or 12.8~${\rm  \mu m}$ [Ne$\textsc{ii}$] emission features, which at these redshifts would pass through the 24~${\rm \mu m}$ passbands; and 4) the high-$z$ sample with $z>1.3$, a range, which   includes the possible peak at $z\sim 2.4$ in which we expect a significant incompleteness since sources could fall out of the detection limits due to the deep silicate absorption features at $\sim$~9.7~${\rm  \mu m}$.
 At $z\sim 1.5, $ these features would appear in $\sim$24~${\rm  \mu {\rm m}}$ passbands. 
For the subsamples divided in this way, we estimated the angular correlation function.
What is more, as a by-product of the photometric redshift estimation we were able to obtain the estimates of the $L_{TIR}$ for the galaxies in the AKARI 24 $\mu$m selected sample. The details of the analysis of the physical parameters for the sources from AKARI NEP Deep Field using CIGALE will be addressed in Malek et. al (in prep) and Buat et al. (in prep).
Here we use the output for general purposes only.

\begin{figure}[!h]
\centering
\includegraphics[width=0.4\textwidth]{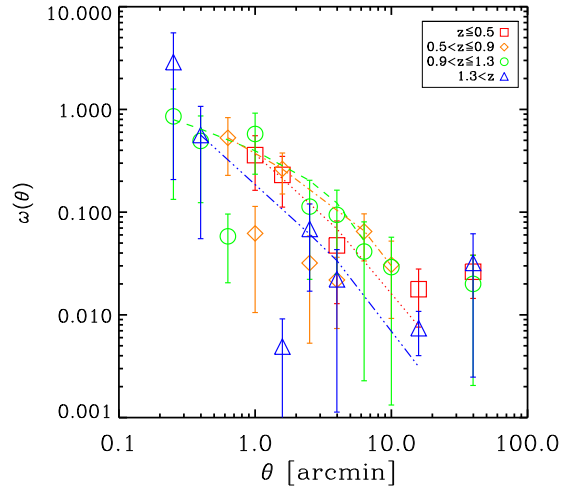} 
\caption{Two-point angular correlation function calculated for the subsamples divided according to the derived photometric redshifts with $\chi^2 <10$. Squares represent the low z sample ($z \leq 0.5$), diamonds  a sample with $0.5<z\leq0.9$, circles  the sample with $0.9< z \leq 1.3,$ and triangles the high z sample with $z> 1.3$.}
\label{hilow1}
\end{figure}

 Fig.~\ref{hilow1} shows the results of this calculation, while the Table~\ref{tab2} lists the derived power-law fitting parameters along with the corresponding correlation lengths, $L_{TIR}$, and bias (see subsection 4.2) for galaxies in each redshift bin.
 Table~\ref{cezrange} summarizes the amount of the catastrophic errors obtained within each considered redshift range.
% The highest redshift bin, due to a high content of CE, is unreliable. 
%Nevertheless, since the amount of spectroscopic redshift measurements on which the calculation of CE is based on is rather small in general, we shall use the high-z subsample for a comparison.
 %Nevertheless, the results obtained for this range may be treated as tentative due to the fact that the CE calculation is based on a very small sample.

%The derived correlation lengths for the {\bf secure sub-sample and for the full catalog} differ dramatically for every redshift range (Table \ref{tab2}).
%This effect is caused by the amount of the CE contained within the {\bf $\chi^2<99$} sample, which is significantly higher than for the $\chi^2<10$ sample (Tab. \ref{cezrange}).
%The result of the too high fraction of CE's contained within a sub-sample results in a dilution of the signal and a significant decrease in the correlation length.
%Substitution of the photometric redshifts with spectroscopic ones does not improve the calculations, unlike in case of the $\chi^2<10$ sample (see Sect. 6). % however the statistics indicate, that the 
%While the slopes  of the high and low samples do not differentiate from one another ($\gamma_{low}\sim 1.76$ and $\gamma_{high}\sim 1.77$), the clustering signal is much stronger for the high-$z$ sample ($A^{high}_{\omega}=13.99\pm0.47 \cdot10^{-3}$).
% Since the calculations based on the $\chi^2<10$ sample have proved to be reliable, we discuss the scientific results only for this sample.
 For the lowest redshift interval ($z\leq 0.5$), which predominantly contains galaxies with total infrared (TIR) luminosities of an order of $10^{10} L_{\odot}$, the derived correlation length is equal to $3.62 \pm 0.79 h^{-1}\mbox{ Mpc} $. 
%The similarity of those values mean that out to redshift $\sim 0.9$ we are dealing with one population of dusty star-forming galaxies, which is also confirmed by the fact that there is no local minima in the redshift distribution between those two bins.
At redshifts $0.5<z\leq0.9,$ we start to observe a LIRG population ($L_{TIR}\sim10^{11} L_{\odot}$), which displays a higher $r_0$: $6.21\pm 0.78  h^{-1} \mbox{ Mpc}$ and mean $L_{TIR}$$\sim1.4\cdot 10^{11}L_{\odot}$.

The higher-$z$ population ($0.9<z\leq 1.3$) displays a correlation length ($r_{0}=5.86\pm 0.69$ $h^{-1}\mbox{ Mpc}$) similar to that of the lower-z population, however, it displays a significantly higher $L_{TIR}:\sim7\cdot 10^{11}L_{\odot}$. 
%This suggests that these galaxies may be related in some way to the previous one.}
%{\bf However the luminosity measurements show that this subsample contains much more luminous galaxies: 
%This clustering dependance on the infrared luminosity ...
 %, possibly of distant ULIRGs.
%However, a detailed description of the IR luminosity and how it influences the clustering is required to draw more substantial conclusions.
The correlation function of galaxies in the $z>1.3$ has a correlation length of $7.23\pm 0.87 \mbox{ Mpc}  \mbox{ } h^{-1}$. %, which exhibit a strong clustering signal at small scales. %has revealed a strong clustering signal at small scales despite the large size of the redshift range.
Even though this sample has the largest redshift dispersion, the clustering signal is not a subject to any substantial dilution. 
 Nevertheless, the amount of the catastrophic errors within this bin is high ($\sim83\%$, see Table~\ref{cezrange}); therefore without any additional follow-up measurements of redshifts to confirm our results, conclusions based on this bin cannot be reliable.
%This fact points towards the conclusion, that at this redshift range we are dealing with a separate, highly clustered population (or populations).
%This could indicate an existence of separate, highly clustered population . 
%However, those results should be treated with caution, since the photometric redshifts are burdened with higher uncertainty for higher redshift derivation.
 
%\begin{figure}[!h]
%\centering
%\includegraphics[width=0.4\textwidth]{fkorzrange_chis70.eps} 
%\caption{Two-point angular correlation function calculated for the subsamples divided according to the derived photometric redshifts with $\chi^2 <70$. Squares represent the low z sample ($z \leq 0.5$), diamonds represent a sample with $0.5<z\leq0.9$, circles the sample with $0.9< z \leq 1.3$ and triangles the high z sample with $z> 1.3$.}
%\label{hilow}
%\end{figure}
\begin{table*}[ht]
\caption{ Clustering of 24 $\mu$m galaxies in AKARI NEP Deep Field.} %of the amount of the CE within the specific redshift ranges on the level of $\chi^2 < 10$ }%and $\chi^2 < 70$.}
\begin{center}
\begin{tabular}{cccccccc}
\hline\hline
$\chi^2$&$z$ range  & ${\rm N_{gal}}$ &  ${\rm \gamma}$ &$ r_0\mbox{ } [h^{-1}\mbox{Mpc }]$&$L_{TIR}$& b\\ \hline 
10&$z\leq0.5$&168& $1.99\pm 0.06$& $3.62\pm0.76$&$10.276\pm0.81$&$0.89\pm0.20$\\
10&$0.5<z\leq0.9$&207&$1.69\pm0.03$& $6.21\pm0.78$&$11.15\pm0.315$& $1.72\pm0.25$\\
10&$0.9<z\leq1.3$&159& $1.57\pm0.10$& $5.86\pm0.69$&$11.84\pm0.36$&$1.91\pm0.21$\\
10&$z>1.3$&77&$1.85\pm0.12$&$7.23\pm0.87$&$12.43\pm0.44$& $3.53\pm0.51$\\\hline\hline
%70&$z\leq0.5$&385&$2.71\pm0.73$&$1.73\pm 0.08$&$2.24\pm 0.92$\\
%70&$0.5<z\leq0.9$&425&$4.08\pm0.99$&$1.72\pm 0.03$&$2.82\pm 0.97$\\
%70&$0.9<z\leq1.3$&221&$5.60\pm 0.83$&$1.73\pm 0.06$&$2.77\pm0.83$\\
%70&$z>1.3$&189&$12.34\pm 0.94$&$1.75\pm 0.11$&$4.94\pm 0.80$\\
%\hline
%$\chi^2$&$z$ range  & ${\rm N_{gal}}$ & $A$ [$\times {\rm 10^{-3}}$] (at ${\rm 1^\circ}$) & ${\rm \gamma}$ &$ r_0\mbox{ } [h^{-1}\mbox{Mpc }]$&{\bf$L_{TIR}}$\\ \hline 
%10&$z\leq0.5$&168&$11.64\pm0.99$&$1.87\pm 0.07$&$5.66\pm0.89$&{\bf$10.276\pm0.81$}\\
%10&$0.5<z\leq0.9$&207&$15.96\pm0.98$&$1.78\pm0.02$&$6.88\pm0.76$&${\bf11.15\pm0.315$}\\
%10&$0.9<z\leq1.3$&159&$17.45\pm2.52$&$1.72\pm0.14$&$5.87\pm0.79$&{\bf$11.84\pm0.36$}\\
%10&$z>1.3$&77&$8.68\pm 1.26$&$1.85\pm0.12$&$7.23\pm0.87$&{\bf$12.43\pm0.44$}\\\hline\hline
\end{tabular}
\end{center}
\label{tab2}
\end{table*}

\begin{table}[!h]
\caption{Comparison of $z_{\rm CIGALE}$ vs $ z_{\rm spec}$ as a function of $\chi^2$. $\eta_{\rm{bin}}$  - percent of CE in the redshift bin, $\eta_{\rm{ tot}}$ - percent of CE in the total matched sample.}
\begin{center}
\begin{tabular}{c c c c c c }
\hline\hline
$\chi^2$ & $z_{\rm range}$& $ N_{\rm gal}$ &  $N_{\rm CE}$ & $\eta_{\rm{bin}}$ [\%]& $\eta_{\rm{ tot}}$ [\%]\\ \hline
10&$z\leq 0.5$&32&0&0.00&0.00\\
10& $0.5<z\leq 0.9$&5&1&20.00&1.92\\
10& $0.9<z\leq 1.3$&6&3&50.00&5.77\\
10& $1.3<z$&6&5&83.33&9.62\\
\hline
 \end{tabular}
 \end{center}
 \label{cezrange}
\end{table}

\subsection{Galaxy bias}

We study the bias $b$ of the 24 $\mu$m selected AKARI galaxies as described in subsection 3.5. We leave a more detailed study using a full Halo Occupation Distribution (HOD, \citealt{hod}) formalism for future work.
\begin{figure}[!h]
\centering
\includegraphics[width=0.4\textwidth]{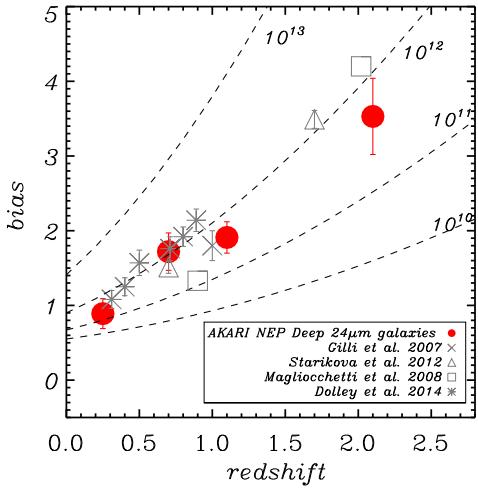} 
%\resizebox{0.75\hsize}{!}{\includegraphics{bM.eps}}
        \caption{ Linear bias as a function of redshift for AKARI photometric redshift samples (filled circles). Dashed curves represent the theoretical linear halo bias evolution of dark matter halos of minimal masses: $10^{10}$, $10^{11}$, $10^{12}$, and $10^{13}$ (from bottom to top). As a reference we show results from the literature: triangles mark results from \citet{starikova}, squares the results from \citet{maglio}, crossesthe results from \citet{gilli}, and  asterisks  from \citet{dolley14}.}  
%The solid lines have been computed
%using the fitting functions of Sheth & Tormen (1999) for the halo mass
%function and bias. The label for each of these lines indicates the minimum halo mass in terms of $log[M_{\odot} h^{-1}]$.}

     \label{haloevo}
\end{figure}

%\subsubsection{Bias evolution}
Once the bias factors have been computed for all AKARI galaxy subsamples (section 3.5), it is possible to trace their evolution with redshift. This procedure can give an insight into what kinds of galaxies the 24 $\mu$m selected samples could correspond to at the current epoch: $z=0$ (\citealt{nusser94}, \citealt{mosc98}). This approach is however simplified: it is based on the assumption that galaxies throughout their evolution do not interact with each other and they are only influenced  by their density field.
We can track the evolution by the relation,
i.e.,\begin{equation}
b(z)= 1+\frac{b(0)-1}{D(z)}
\label{bz1}
,\end{equation}
where ${\rm D(z)}$ is given by eq.~\ref{dz}, and $b(0)$ is a value of bias at $z=0$. 

%In order to validate the linear halo bias we have used the fitting functions for halo mass function and bias (following \citealt{sheth}).
In Fig.~\ref{haloevo} we show the linear bias evolution derived from \citet{sheth} formalism for varying minimum dark matter halo (DMH) mass thresholds.
The minimum masses of DMH for galaxies within redshift intervals of $z<0.5$ and $0.9<z\le1.3 $ are located between $10^{11}M_{\odot}h^{-1}<M_{h}<10^{12}M_{\odot}h^{-1}$. For $0.5<z\le0.9$ interval, galaxies are expected to have a slightly higher minimum mass than other considered samples: $10^{12}M_{\odot}h^{-1}\ge M_{h}$. However, since values of bias parameters for the intervals of  $0.5<z\le0.9$ ($b=1.72\pm0.25$) and $0.9<z\le1.3 $ ($b=1.91\pm 0.21$) are in agreement within their error bars, this result for the lower $z$ sample could be misleading.
 Nevertheless, if it is a true effect, it could indicate that IR galaxies with higher $L_{TIR}$ do not necessarily reside in more massive DMHs, as in the case of redshift intervals of $0.9<z\le1.3 $ (with $L_{TIR}\sim7\cdot10^{11}L_{\odot}$) and  $0.5<z\le0.9$ (with $L_{TIR}\sim1.4\cdot 10^{11}L_{\odot}$). This idea is consistent with the work of \citet{goto05}, where it has been reported (for the IR galaxies observed by IRAS) that more IR luminous galaxies tend to have smaller local density. They argue that luminous infrared galaxies with moderate total infrared luminosities ($10^{10.5}L_{\odot}<L_{TIR}<10^{11}L_{\odot}$) exist in  higher density regions than IR galaxies with $10^{11}L_{\odot}<L_{TIR}<10^{12}L_{\odot}$.
Therefore there must exist a discrepancy in the origin of these two potentially different populations. If LIRGs are created through a merger and/or interaction of several galaxies (e.g., \citealt{taniguchi}, \citealt{borne00}), then their local environment would naturally be lower. %Results of \citet{goto05} and of this work seem to confirm the fact,
Then, the higher minimal DMH of the AKARIs' $0.5<z\le0.9$ could mean that within this interval we still observe a substantial fraction of normal spiral galaxies in addition to interacting systems.
It has been reported that up to $z\sim0.5-0.7$ more than a half of the LIRG population is composed of normal spiral galaxies (\citealt{bell05}; \citealt{melbourne05}; \citealt{elbaz07}). Beyond those redshifts, mergers start to enhance the $L_{TIR}$ and deplete the environment.
However, a relatively large $r_{0}$ of AKARIs' galaxies is contradictory to local measurements of clustering lengths of star-forming spiral galaxies ($r_{0}\sim 4$ $h^{-1} {\rm Mpc}$; e.g.,  \citealt{coil04}; \citealt{meneux06}; \citealt{coil08}; \citealt{heinis}). On the other hand, AKARI NEP Deep survey is flux limited, therefore, at larger redshifts it preferentially and intrinsically detects  more luminous galaxies, which in turn are more strongly clustered. This could result in an increase of the measured $r_{0}$ in this redshift range.
This issue remains undetermined and a more detailed analysis of those galaxies is needed to determine their true nature, which is a subject of future research.

With known $b(z)$, we can trace the evolution of $\xi_{g(r=8,z)}$, and therefore the evolution of the $r_{0}(z)$, by inverting Eq.~\ref{biaspar}. This procedure can lead to the determination of a possible fate of each considered population.

\begin{figure}[!h]
\centering
\includegraphics[width=0.4\textwidth]{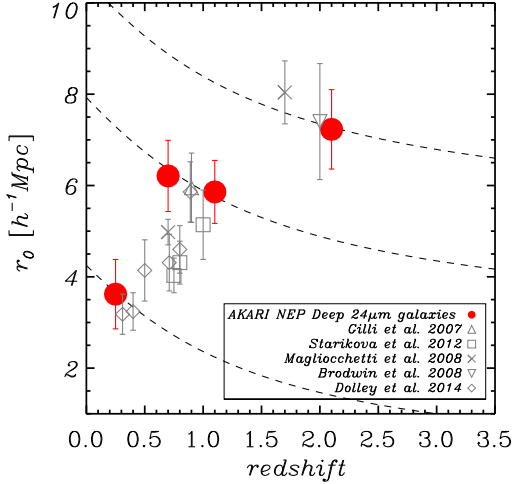} 
%\resizebox{0.75\hsize}{!}{\includegraphics{b0.eps}}
        \caption{ Correlation length of AKARI NEP Deep galaxies (red filled circles) compared with the evolutionary tracks as expected from the linear scenario (dashed lines). Triangles denote results from \citet{gilli}, squares from \citet{starikova}, crosses from \citet{maglio}, reversed triangles from \citet{brodwin}, and diamonds from \citet{dolley14}. }
     \label{revol}
\end{figure}

 %, since bias parameter relates the .
%For this computation we will use a constant slope value, derived for the full catalog ($\gamma=1.79$), and
 The results are shown in Fig.~\ref{revol}.
AKARI 24 $\mu$m galaxies at $z_{med}=0.25$ exhibit a steep correlation function slope ($\gamma\sim2$) and a correlation length of $3.62\pm0.76$ $h^{-1} {\rm Mpc}$.
 Their evolution with time could result in reaching $r_{0} \sim 4 $ $h^{-1} {\rm Mpc}$ at the present day.
%Since this redshift interval may be considered as a measurement for the local Universe, the evolution is not as pronounced as for other redshift intervals.
Therefore, according to those measurements, the majority of the AKARI galaxies at low-z are expected to fall into this category.
%, that elliptical galaxies with $L<L_{*}$ exhibit such strong clustering (e.g. \citealt{heinis}, \citealt{zehavi02}, \citealt{madgwick}) as well as similar steep slope of the correlation function ($\gamma\sim1.9$), which could indicate that some of the infra-red galaxies are their progenitors. 

The galaxies within redshift ranges of $0.5<z\le0.9$ and $0.9<z\le1.3$ could evolve into a very strongly clustered population of galaxies, with $r_{0}\sim$ 8-8.5 $h^{-1} {\rm Mpc}$.
In the local Universe, early type galaxies with $L\sim L_{*}$ have been measured to have similar correlation lengths: $r_{0}\sim 8$ $h^{-1} {\rm Mpc}$ (\citealt{guzzo97}). This means that those objects could have evolved from the population of LIRGs at redshifts $\sim 0.9$. However, the slope of the clustering of AKARI galaxies ($\gamma\sim1.65$) is lower than that measured for the local ellipticals ($\gamma\sim2$). %In the work of \citet{buda03} it was shown that 

%data together with results from the literature. 

\subsection{Comparison with previous studies}
%\begin{figure}[!h]
%\centering
%\includegraphics[width=0.4\textwidth]{r0zcomp.eps} 
%\caption{Comparison of the clustering lengths $r_{0}$ of the angular correlation function $\omega(\theta)$ from the literature as a function of redshift. This work (open circles; only results obtained from the subsample with $\chi^2<10$ are shown) is confronted with: sample of dust obscured galaxies (filled squares) from \citealt{brodwin}; high and low redshift samples from \citealt{starikova} and \citealt{maglio} (triangles and filled diamonds, respectively); LIRGs from \citealt{gilli}; sub-millimetre galaxies (SMG) derived by \citealt{blain04} (marked by the horizontal dotted lines representing the $1 \sigma$ region).}
%\label{r0}
%\end{figure}
 Figures~\ref{haloevo} and \ref{revol} present a comparison of our derived spatial correlation lengths and biases for the redshift divided samples with the results from the literature. 
%Here, only the results for the $\chi^2<10$ subsample will be considered, since the full catalog with $\chi^2<99$ is strongly affected by the CE of the redshift estimation, and therefore the results are unreliable.

As mentioned in Section 1, several previous studies have approached the task of analyzing the 24 ${\rm  \mu \mbox{m}}$ selected galaxies by dividing the samples based on their redshift measurements. 
 A direct comparison with previous studies is not a straightforward procedure because of the  varying survey sizes and depths, different methods used to select the galaxy samples, and the difference in the considered redshift ranges.
 We present the first non-{\it Spitzer} based measurement of clustering of 24-$\mu$m selected sources.
%, which provides a complementary look into the nature of such galaxies. 
Table~\ref{tabb22} shows a comparison of the previously investigated surveys of 24 $\mu$m selected galaxies throughout the literature.
%AKARI data can provide an independent and a complementary insight on the topic, having a deeper observational limit (out of the photometric surveys considered here) and higher surface density of objects.

%\begin{savenotes}
\begin{table*}[ht]
\caption{Summary and comparison of the previous 24 $\mu$m selected galaxy surveys.
The work of Gilli \& Daddi (2007) is based on spectroscopic measurements in contrast with other studies (based on photometric measurements). \ The parameter $\rho$ stands for surface density.}
\label{prop}
\begin{center}
\begin{tabular}{c|c|c|c|c|c|c}
\hline\hline

 &depth [$\mu$Jy]&$\rho^{\footnotemark[1]}$ [${\rm N}_{{\rm gal}}/deg^{2}$] &size [$deg^{2}$]&probed $z$&$\gamma$&$r_{0} [\mbox{ }h^{-1}\mbox{Mpc}]$\\\hline
Gilli \& Daddi 2007 &20&13060&0.1&$\sim1$&$1.5\pm0.1$&$4.0\pm0.4$\\
Magliocchetti et al. 2008& 400&1490&0.7&[0.6-1.2] \& $<1.6$&--&$5.95_{-1.26}^{+1.10}$\& $11.13_{-2.38}^{+2.03}$\\
Starikova et al. 2012&310&410&49&$\sim0.7$ \& $\sim1.7$&1.8&$4.98\pm0.28$\&$8.04\pm0.69$\\
Dolley et al. 2014& 223&2678&8.42&$0.2<z<1.0$&$1.8-2.13$&$3-6$\\\hline
AKARI&150&3350&0.4&$[0:3)$&$1.79\pm0.04$&$4.93\pm0.49$
\\\hline\hline

\end{tabular}
\end{center}
\label{tabb22}
\end{table*}

%\end{savenotes}
%\footnotetext{based on spectroscopic measurements}
%Out of all photometry-based anayses, AKARI observations are the deepest. }
 It is worth noting that the source selection, which we have relied upon for this analysis, was based on the rough spectral shape of the considered sources. This excludes the possibility of mixing AGN dominated galaxies, in contrast with samples chosen without this kind of  criterion. 
%Moreover, such an approach to source classification presents very little statistical biases, and does not rely on availability of optical counterparts or on redshift cuts.
%Moreover, the SVM method forces very few constraints on the selection. We have worked on a full range of redshifts, which allowed for an unbiased emergence of different populations of sources.

 Measurements of the correlation length of local ($0.3<z<0.6$) blue galaxies have been reported in several previous studies to be on the order of $\sim 3.7$ $h^{-1} {\rm Mpc}$ (e.g., \citealt{coil04}). This was found to be in agreement with the measurements of the rest-frame selected UV galaxies (see \citealt{milliard}, \citealt{heinis}). Since this value is consistent with that measured for the lowest-z AKARI galaxy subsample ($r_0=3.62\pm0.76$ $h^{-1} {\rm Mpc}$), it could mean that it is mainly composed of the local, normal SF galaxies.  
The study by \citet{masci} for 24~$\mu$m selected galaxies in SWIRE survey found similar values of  $r_0$: $3.32\pm0.19$~$h^{-1} {\rm Mpc}$ with the steep slope $\gamma\sim2$.

For the sample of $\sim$~1300 objects ($S_{24 \mu \mbox{m}}>20 \mbox{ }\mu \mbox{Jy}$) with an average $L_{\rm IR}\sim4\cdot 10^{10}L_{\odot}$ within the GOODS fields with $z_{med}\sim0.8$, \citealt{gilli} have found the correlation length equal to $ r_0=4.0\pm0.4$ $h^{-1} {\rm Mpc}$.
They have found that for LIRGs ($L_{\rm IR}>10^{11}L_{\odot}$)  $ r_0$ rises up to $5.14\pm0.76$ $h^{-1} {\rm Mpc}$.
%The bias parameter measurements from both studies agree as well, with 
These values are in agreement with that obtained for the LIRGS found in AKARI NEP-Deep survey.
% on redshift ranges of $z\leq0.9$. %: ${\rm r_0}=5.91\pm0.66$ ${\rm Mpc}$ ${\rm h^{-1}}  $.

%This research also revealed that the correlation length increases with the increase of the IR luminosity reaching the value of $5.14 \pm 0.76 $ ${\rm Mpc}$ ${\rm h^{-1}}  $ for the brightest ones ($L_{IR}>10^{11} L_{\odot}$). %This value is consistent with the value from our work of $r_0=5.91 \pm 0.66 $ ${\rm Mpc}$ ${\rm h^{-1}}  $, obtained for the similar redshift ranges ($z_{median}\sim 0.5$ and $z_{median}\sim 0.7$ ).
% Since, according to our redshift estimates, AKARI galaxy sample consists mainly of the lower-z galaxies 

%%%%%%%%%%%%%%%%%%%%%%%%%%%%%MASCI%%%%%%%%%%%%%%%%%%%%%%%%%%%%%%%

%Moreover, our resultant correlation length is larger than the one obtained from the IRAS and ISO study at these wavelengths performed by \citealt{masci} for $\sim 24000$ sources brighter than $S_{24 \mu \mbox{m}}=250 \mu \mbox{Jy}$ with median redshift $\sim 0.8$.
% The reported correlation length is equal to ${\rm r_0=}3.32\pm0.19$ ${\rm Mpc}$  ${\rm h^{-1}}$.
%If we were to compare this value with the one derived for the $250 \mu$m-limited catalog, the median redshift for this sample is $\sim 1.5$, so the volumes probed by AKARI NEP galaxies is larger.
%These results are consistent with the ones obtained by ISO and IRAS in the MIR.
%The reason behind these discrepancies may be that...??? ojojoj 

%%%%%%%%%%%%%%%%%%%%%%%%%%%%%%%%%%%%%%%%%%%%%%%%%%%
%%%%%%%%%%%%%%%%%%%%%%%%%%%%MAGLIOCCHETTI%%%%%%%%%%%%%%%%%%%%%%%%%%%%%%%%%%%%%%
The analysis performed by \citealt{maglio} for 1041 galaxies  brighter than $S_{24 \mu {\rm m}}=400$ $\mu \mbox{Jy}$ resulted in obtaining the correlation lengths for low-$z$ ($z_{\rm mean}\sim0.8$) and high-$z$ ($z_{\rm mean}\sim2$) sources to be equal to $ r_0=5.95^{+1.10}_{-1.26}$ $h^{-1}{\rm Mpc}$  and $ r_0=11.13^{+2.03}_{-2.38}$ $h^{-1} {\rm Mpc}$, respectively.
The value obtained for the low-$z$ sample, $z=[0.6-1.2]$, is comparable with that derived in this work ($ r_0=5.86\pm0.69$ $h^{-1}{\rm Mpc}$) for $0.9<z\leq1.3$ range. 
%{\bf The derived bias parameter for the galaxies with $z\sim1$ are comparable ($b_{{\rm AKARI}}\sim 2.3$ vs $b_{{\rm Magliocchetti}\sim1.7}$),
This evidence for both studies at these redshifts points to the fact that they trace a population of galaxies that could evolve into present day $ r_0\sim8$ $h^{-1}{\rm Mpc}$ objects (possibly ellipticals with $L\sim L_{*}$).
%However AKARI galaxies exhibit lower DMH masses ($M_{min}\sim0.8\cdot 10^{11}M_{\odot}$ vs $M_{min}\sim 10^{12}_{Sun}$).  
%This difference could arise from the fact, that the selection procedures were drastically different, where in case of AKARI 

%The high-$z$ sample
%The discrepancy between those values originates in the fact, that despite probing similar redshifts, AKARI sample reaches into deeper fluxes ($\sim{\rm 150 \mu Jy}$). 
%{\bf not sure if this makes sense and whether im allowed to do this:}\\
%Moreover, if we take into the account the correlation length derived for the ${\rm 400 \mu Jy}$ limited sample (with $z_{{\rm mean}}\sim 1.5$), the AKARI correlation length $ r_0=8.79\pm 0.49$  $h^{-1}{\rm Mpc}$ is consistent with the value derived for the high-$z$ sample within the errorbars. 
%Even though the obtained values are different from the ones obtained by our analysis, this study probes galaxies existing much deeper our sample.
%%masakrycznie sie roznia!
%%%%%%%%%%%%%%%%%%%%%%%%%%%%%%%%%%%%%%%%%%%%%%%%%%%%%%%%%%
%%%%%%%%%%%%%%%%%%%%%%%%%STARIKOVA%%%%%%%%%%%%%%%%%%%%%%%%%%%%%%%%%%%%%%%%

The study by \citealt{starikova}, based on Spitzer Wide-Area Survey within the SWIRE Lockman Hole field for more than 20000 objects brighter than $S_{24 \mu {\rm m}}>310$ $\mu \mbox{Jy,}$ has shown that the correlation length for the low- ($z_{mean}\sim0.7$) and high-$z$ ($z_{\rm mean}\sim1.7$) samples are $ r_0=4.98\pm0.28 $ ${\rm Mpc}$ ${\rm h^{-1}} $ and ${\rm r_0=}8.04 \pm 0.69 $  $ {\rm Mpc}$ ${\rm  h^{-1}}$, respectively. 
Our derived values  for $0.5<z\leq 0.9$ are slightly higher ($r_0=6.21\pm0.78 $ $h^{-1} {\rm Mpc}$). %, however the considered redshift range differs significantly.
 % the redshifts intervals that we compare are not the same and therefore each interval could  contain different mixtures of populations.
 %The study by \citet{starikova} covers an area of 49 deg$^{2}$ and contains more than 20000 objects, which is equivalent to the surface density of approximtely 410 galaxies per square degree. In contrast AKARI NEP Deep survey coveres a field of $\sim 0.4$deg$^{2}$, where 1339 galaxies are considered, which translates to a surface density of approximately 3350 galaxies per square degree (see Table \ref{tabb22}).\\
%\\
%The study by \citet{starikova} reaches very deep into space despite very large area coverage. AKARI area is much smaller and reaches slightly deeper into space, which results in detecting objects intrinsically brighter albeit at larger distances. 
%The redshift estimation presented by \citet{starikova} is based on a combination of applying a color-magnitude relation
%In the study of \citet{starikova} the samples were divided into high and low redshift subsamples 
%based on a color-magnitude relation and its dependance on redshift of sources form a GOODS survey (referencja) with similar selection criteria. In this work we have performed a photometric redshift estimation based on the template fitting.
The divergence of the results could arise because the widths of redshift intervals that are being compared are different. The low-redshift interval of Starikova et al. covers a range from $\sim 0$ to 2, with the mean $z\sim 0.7$. 
In AKARI's range for $z_{median}\sim 0.7,$ we have not included the galaxies at $z$ lower than 0.5.
%Therefore the dissimilarities of the results between this work and the work of \citet{starikova} may be due to the difference between the derivation of the redshift distribution.
% therefore direct comparison is not straightforward, and should be treated with caution.

 Furthermore, the recent study by \citet{dolley14} from the Spitzer observations of the Bo\"otes field of over 22~000 24~$\mu$m galaxies up to  $z\sim1$ shows correlation lengths  varying with redshift, from $r_{0}\sim 3$ to $r_{0}\sim 6$ $h^{-1}{\rm Mpc}$. Subsamples at $z_{median}\sim0.3$ with $r_{0}=3.18\pm 0.44$ $h^{-1}{\rm Mpc}$ and $z_{median}\sim 0.9$ with $r_{0}=5.86\pm 0.66$ $h^{-1}{\rm Mpc}$ show a good agreement with our results   at similar median redshifts: $r_{0}= 3.62\pm0.76$  $h^{-1}{\rm Mpc}$ at $z_{median}\sim 0.25$ and $r_{0}= 6.21\pm0.78$ $h^{-1}{\rm Mpc}$ at $z_{median}\sim 0.7$.
 It is also worth noting that the subsample of AKARI galaxies with the highest $z$ indicates a correlation length value ($ r_{0}=7.23\pm0.87$ $h^{-1}{\rm Mpc}$ ), which is consistent with the work of \citet{brodwin}. Their investigation was focused on clustering analysis of $\sim 2600$ dust obscured galaxies (DOG) with spectroscopic redshifts located at $1.5<z<2.5$. A detection of strong clustering with $r_{0}=7.4^{+1.27}_{-0.84}$ $h^{-1}{\rm Mpc}$ for a ${\rm S_{24 \mu m}}>$~300~$\mu$Jy flux cut was reported. 
 %with whom our results  for the highest redshift subsample are comparable. 
However, these values from the AKARI analysis were derived from small number statistics, and therefore, these results should be treated with caution.

%What is more, our results are consistent with the lower limits for the correlation length derived for the submilimetere galaxies (SMG) by falling into the $1\sigma$ area derived by \citealt{blain04}.
%On the other hand, \citet{coo} reported clustering measurements of sources detected down to 30 mJy revealed the correlation lengths of $ r_0=3.15\pm0.35$ $h^{-1} {\rm Mpc}$ and $ r_0=4.41\pm0.49$ $h^{-1}{\rm Mpc}$ with the mean redshift ranges of $z\sim 2.1$ and $\sim 2.6$, at 250 ${\rm \mu m}$ and 500 ${\rm \mu m}$, respectively, which means that the 24 ${\rm \mu m}$ population(s) of AKARI sources cannot be related to them.

%%%%%%%%% Flux limit tutaj to jest po prostu Flux dla najciemniejszego obiektu w danej probce.
\section{ Summary and conclusions}

%A more indepth analysis of the

We have presented a clustering analysis of the $S_{24\mu {\rm m}}$ selected AKARI NEP Deep field galaxies down to a flux limit of 150 $\mu$Jy.
This is the first study of these kinds of sources that are not  based on Spitzer telescope measurements  and features a source selection that does not rely on any auxiliary observations from other wavelength intervals.

After performing the sample selection through the use of the SVM classifiers trained on the infrared color information, we  used the CIGALE code to estimate the photometric redshifts for the chosen galaxies, and we were able to establish a redshift estimate for 96.11 \% of the sample. 
The resultant redshift distribution revealed three peaks: the primary one located at $z\sim 0.6$, which can be attributed to either the evolution in the luminosity function or to the increase of the detection volume; a secondary peak located at $z\sim 1.2$ possibly related to the $12.7$ and 12.8 $ \mu $m PAH emission lines passing through the 24 $\mu $m passband at these redshifts; and an indication of a possible third peak at $z\sim2.4$. Even though AKARI's high-z peak is too tentative to draw any firm conclusions about any possible population of galaxies, several models predict its existence and attribute it to the sources dominated either by AGNs or by a PAH emission feature at 8 $ \mu $m.

For the calibration of those measurements, we have compared them with the spectroscopic redshifts for counterparts available from Murata et al. catalog (149 objects) and estimated that, depending on the goodness of fit value, we encounter a 18.75 \% rate of CE for $\chi^2 <10$ and 35.22~\% for $\chi^2<70$. 
Therefore, the following clustering analysis was performed only on a subsample of galaxies with the most secure redshift estimation.
%However, since the difference in the estimates of $\chi^2$ depends on the distance of the outlying data points from the theoretical SED model, we decided to perform the analysis for two samples of galaxies; for the ones with photometric redshift estimation on the levels of $\chi^2<10$ and $\chi^2<70$.
%The galaxies with $\chi^2$ higher than 80 were excluded from any calculations.
To obtain the most reliable results, we  substituted the photometric redshift measurements for the objects possessing the spectroscopic redshift measurements. 
With a set of galaxies prepared in the way described above, we  examined the clustering properties of the derived samples with respect to their redshift distribution.% and their dependence on brightness limits.

Using a power-law approximation to derive correlation function together with the estimated photometric redshift information, we obtained the spatial correlation length $r_0$. For the full sample, the derived value equals $r_0=4.93\pm0.49$ $h^{-1}{\rm Mpc}$, which is consistent with the previous studies conducted for the same wavelengths.
 For the specific redshift intervals we have found the following facts: 
\begin{enumerate}
\item{For $z\leq 0.5$ $r_{0}=3.62 \pm 0.79$ $h^{-1}\mbox{ Mpc}$,  $L_{TIR}$ $\sim10^{10}L_{\odot}$. This indicates that this redshift interval is mostly composed of normal SF galaxies. Their bias parameter is equal to $0.93\pm0.11$ and their estimated minimal DMH mass is $M_{h}>10^{11}M_{\odot}$. }
\item{For $0.5<z\leq0.9$ and $0.9<z\leq 1.3$:}
\begin{itemize}
\item{correlation lengths are $6.21\pm 0.78$ $h^{-1}\mbox{ Mpc}$ and $5.86\pm 0.69$ $h^{-1}\mbox{ Mpc} $, respectively ;}
%\item{infrared luminosities of galaxies in both bins are lower than $\sim8\cdot10^{11} L_{\odot}$;}
\item{galaxies exhibit very similar, relatively high values of bias parameter  ($\sim 1.7-1.9$). 
 Those values are consistent with the linear bias derived in work of \citet{lagache07}, which was measured from Cosmic Far-Infrared Background anisotropies at 160 $\mu$m ($b=1.74\pm0.16$). Both measurements probe similar redshift range ($z_{Lagache}\sim 1$).}

\item{the sample at $0.9<z\leq 1.3$, composed of brighter galaxies ($L_{TIR} \sim 10^{11.84} L_{\odot}$), seems to be residing in lower minimal mass DMH ($M_{h}<10^{12}M_{\odot}h^{-1}$) than the lower luminosity (with $ L_{TIR} \sim 10^{11.15} L_{\odot}$ and $M_{h}>10^{12}M_{\odot}h^{-1}$) galaxies at $0.5<z\leq0.9$. This could indicate that brighter infrared galaxies do not necessarily reside in more massive halos. 
This means that despite  similar clustering properties, we are dealing with two different populations of star-forming galaxies. The redshift distribution shows two distinct peaks (at $\sim0.6$ and $\sim 1.2$): the primary peak in the redshift distribution at $\sim 0.6$ could contain a mix of both normal star-forming galaxies and LIRGs. The dip in the distribution between the peaks at $\sim 0.9$  could mean that normal galaxies slowly fade out from the field of view to reveal a more prominent LIRG population at redshifts $\sim0.9$ and higher. }
\end{itemize}

\begin{itemize}
\item{Extrapolating the AKARI data to the present day epoch has allowed us to investigate the approximate descendants of each galaxy subsample. Working under the assumption that galaxy interactions do not affect the $\sigma_{m}$ value at scales as large as 8 Mpc, 24~$\mu$m galaxies at redshifts $0.5<z<1.3$ could evolve into current galaxies with $r_{0}\sim8$ $h^{-1} {\rm Mpc}$, a value measured for ellipticals with $L\sim L_{*}.$}
\end{itemize}

\item{Galaxies at $z>1.3$ display a strong clustering signal at small scales despite the large size of the redshift range.
This could indicate an existence of separate, highly clustered population(s). 
However, because of  small number statistics burdened with a large number  of CE, these results remain unconfirmed.}
\end{enumerate}

\begin{acknowledgements}
 We thank the anonymous referee for helpful suggestions and constructive criticisms, which greatly helped to improve the manuscript.
This work is based on observations with AKARI, a JAXA project with the participation of ESA.
%AS, AP and KM have been supported by the National Science Centre grants UMO-2015/16/S/ST9/00438, UMO-2012/07/B/ST9/04425 and UMO-2013/09/D/ST9/04030.
AP and KM have been supported by the National Science Centre grants, UMO-2012/07/B/ST9/04425 and UMO-2013/09/D/ST9/04030.
This research was partially supported by the project POLISH-SWISS ASTRO PROJECT 
cofinanced by a grant from Switzerland through the Swiss Contribution 
to the enlarged European Union.
TTT has been supported by 
the Grant-in- Aid for the Scientific Research Fund (23340046,
and 24111707), for the Global COE Program Request
for Fundamental Principles in the Universe: from Particles
to the Solar System and the Cosmos commissioned by the
Ministry of Education, Culture, Sports, Science and Technology (MEXT) of Japan, and for the JSPS Strategic Young Researcher Overseas Visits Program for Accelerating Brain Circulation, “Construction of a Global 7 Platform for the Study of Sustainable Humanosphere.”

\end{acknowledgements}

\bibliographystyle{aa}
\bibliography{solarz.bib}

\begin{thebibliography}{100}
\expandafter\ifx\csname natexlab\endcsname\relax\def\natexlab#1{#1}\fi

\bibitem[{{Bardeen}(1986)}]{bard86}
{Bardeen}, J.~M. 1986, {Galaxy formation in an Omega = 1 cold dark matter
  universe}, ed. E.~W. {Kolb}, M.~S. {Turner}, D.~{Lindley}, K.~{Olive}, \&
  D.~{Seckel}, 212--217

\bibitem[{{Bardeen} {et~al.}(1986){Bardeen}, {Bond}, {Kaiser}, \&
  {Szalay}}]{bardeen}
{Bardeen}, J.~M., {Bond}, J.~R., {Kaiser}, N., \& {Szalay}, A.~S. 1986, \apj,
  304, 15

\bibitem[{{Barrow} {et~al.}(1984){Barrow}, {Bhavsar}, \& {Sonoda}}]{boot}
{Barrow}, J.~D., {Bhavsar}, S.~P., \& {Sonoda}, D.~H. 1984, \mnras, 210, 19P

\bibitem[{{Bell} {et~al.}(2005){Bell}, {Papovich}, {Wolf}, {Le Floc'h},
  {Caldwell}, {Barden}, {Egami}, {McIntosh}, {Meisenheimer},
  {P{\'e}rez-Gonz{\'a}lez}, {Rieke}, {Rieke}, {Rigby}, \& {Rix}}]{bell05}
{Bell}, E.~F., {Papovich}, C., {Wolf}, C., {et~al.} 2005, \apj, 625, 23

\bibitem[{{Bertin} \& {Arnouts}(1996)}]{sex}
{Bertin}, E. \& {Arnouts}, S. 1996, \aaps, 117, 393

\bibitem[{{Boissier}(2013)}]{samuel}
{Boissier}, S. 2013, {Star Formation in Galaxies}, ed. T.~D. {Oswalt} \& W.~C.
  {Keel}, 141

\bibitem[{{Borne} {et~al.}(2000){Borne}, {Bushouse}, {Lucas}, \&
  {Colina}}]{borne00}
{Borne}, K.~D., {Bushouse}, H., {Lucas}, R.~A., \& {Colina}, L. 2000, \apjl,
  529, L77

\bibitem[{{Boselli} {et~al.}(2014){Boselli}, {Voyer}, {Boissier}, {Cucciati},
  {Consolandi}, {Cortese}, {Fumagalli}, {Gavazzi}, {Heinis}, {Roehlly}, \&
  {Toloba}}]{boselli}
{Boselli}, A., {Voyer}, E., {Boissier}, S., {et~al.} 2014, \aap, 570, A69

\bibitem[{{Brodwin} {et~al.}(2008){Brodwin}, {Dey}, {Brown}, {Pope}, {Armus},
  {Bussmann}, {Desai}, {Jannuzi}, \& {Le Floc'h}}]{brodwin}
{Brodwin}, M., {Dey}, A., {Brown}, M.~J.~I., {et~al.} 2008, \apjl, 687, L65

\bibitem[{{Buat} {et~al.}(2007){Buat}, {Takeuchi}, {Iglesias-P{\'a}ramo}, {Xu},
  {Burgarella}, {Boselli}, {Barlow}, {Bianchi}, {Donas}, {Forster}, {Friedman},
  {Heckman}, {Lee}, {Madore}, {Martin}, {Milliard}, {Morissey}, {Neff}, {Rich},
  {Schiminovich}, {Seibert}, {Small}, {Szalay}, {Welsh}, {Wyder}, \&
  {Yi}}]{ver07}
{Buat}, V., {Takeuchi}, T.~T., {Iglesias-P{\'a}ramo}, J., {et~al.} 2007, \apjs,
  173, 404

\bibitem[{{Calzetti} {et~al.}(2000){Calzetti}, {Armus}, {Bohlin}, {Kinney},
  {Koornneef}, \& {Storchi-Bergmann}}]{calzetti00}
{Calzetti}, D., {Armus}, L., {Bohlin}, R.~C., {et~al.} 2000, \apj, 533, 682

\bibitem[{{Caputi} {et~al.}(2006){Caputi}, {Dole}, {Lagache}, \&
  {Puget}}]{caputi}
{Caputi}, K.~I., {Dole}, H., {Lagache}, G., \& {Puget}, J.~. 2006, ArXiv
  e-prints: 0604236v1

\bibitem[{{Carroll} {et~al.}(1992){Carroll}, {Press}, \& {Turner}}]{carroll92}
{Carroll}, S.~M., {Press}, W.~H., \& {Turner}, E.~L. 1992, \araa, 30, 499

\bibitem[{{Choi} {et~al.}(2015){Choi}, {Ostriker}, {Naab}, {Oser}, \&
  {Moster}}]{choi}
{Choi}, E., {Ostriker}, J.~P., {Naab}, T., {Oser}, L., \& {Moster}, B.~P. 2015,
  \mnras, 449, 4105

\bibitem[{{Coil} {et~al.}(2008){Coil}, {Newman}, {Croton}, {Cooper}, {Davis},
  {Faber}, {Gerke}, {Koo}, {Padmanabhan}, {Wechsler}, \& {Weiner}}]{coil08}
{Coil}, A.~L., {Newman}, J.~A., {Croton}, D., {et~al.} 2008, \apj, 672, 153

\bibitem[{{Coil} {et~al.}(2004){Coil}, {Newman}, {Kaiser}, {Davis}, {Ma},
  {Kocevski}, \& {Koo}}]{coil04}
{Coil}, A.~L., {Newman}, J.~A., {Kaiser}, N., {et~al.} 2004, \apj, 617, 765

\bibitem[{Cristianini \& Shawe-Taylor(2000)}]{crist}
Cristianini, N. \& Shawe-Taylor, J. 2000, An introduction to Support Vector
  Machines (Cambridge University Press)

\bibitem[{{Dale} \& {Helou}(2002)}]{dale02}
{Dale}, D.~A. \& {Helou}, G. 2002, \apj, 576, 159

\bibitem[{{Davis} \& {Peebles}(1983)}]{dp}
{Davis}, M. \& {Peebles}, P.~J.~E. 1983, \apj, 267, 465

\bibitem[{{Desai} {et~al.}(2008){Desai}, {Soifer}, {Dey}, {Jannuzi}, {Le
  Floc'h}, {Bian}, {Brand}, {Brown}, {Armus}, {Weedman}, {Cool}, {Stern}, \&
  {Brodwin}}]{desai}
{Desai}, V., {Soifer}, B.~T., {Dey}, A., {et~al.} 2008, \apj, 679, 1204

\bibitem[{{Dole} {et~al.}(2004){Dole}, {Le Floc'h}, {P{\'e}rez-Gonz{\'a}lez},
  {Papovich}, {Egami}, {Lagache}, {Alonso-Herrero}, {Engelbracht}, {Gordon},
  {Hines}, {Krause}, {Misselt}, {Morrison}, {Rieke}, {Rieke}, {Rigby}, {Young},
  {Bai}, {Blaylock}, {Neugebauer}, {Beichman}, {Frayer}, {Mould}, \&
  {Richards}}]{dole}
{Dole}, H., {Le Floc'h}, E., {P{\'e}rez-Gonz{\'a}lez}, P.~G., {et~al.} 2004,
  \apjs, 154, 87

\bibitem[{{Dolley} {et~al.}(2014){Dolley}, {Brown}, {Weiner}, {Brodwin},
  {Kochanek}, {Pimbblet}, {Palamara}, {Jannuzi}, {Dey}, {Atlee}, \&
  {Beare}}]{dolley14}
{Dolley}, T., {Brown}, M.~J.~I., {Weiner}, B.~J., {et~al.} 2014, \apj, 797, 125

\bibitem[{{Efstathiou} {et~al.}(1991){Efstathiou}, {Bernstein}, {Tyson},
  {Katz}, \& {Guhathakurta}}]{efstat}
{Efstathiou}, G., {Bernstein}, G., {Tyson}, J.~A., {Katz}, N., \&
  {Guhathakurta}, P. 1991, \apjl, 380, L47

\bibitem[{{Elbaz} \& {Cesarsky}(2003)}]{elbaz}
{Elbaz}, D. \& {Cesarsky}, C.~J. 2003, Science, 300, 270

\bibitem[{{Elbaz} {et~al.}(2007){Elbaz}, {Daddi}, {Le Borgne}, {Dickinson},
  {Alexander}, {Chary}, {Starck}, {Brandt}, {Kitzbichler}, {MacDonald},
  {Nonino}, {Popesso}, {Stern}, \& {Vanzella}}]{elbaz07}
{Elbaz}, D., {Daddi}, E., {Le Borgne}, D., {et~al.} 2007, \aap, 468, 33

\bibitem[{{Faber} {et~al.}(2003){Faber}, {Phillips}, {Kibrick}, {Alcott},
  {Allen}, {Burrous}, {Cantrall}, {Clarke}, {Coil}, {Cowley}, {Davis}, {Deich},
  {Dietsch}, {Gilmore}, {Harper}, {Hilyard}, {Lewis}, {McVeigh}, {Newman},
  {Osborne}, {Schiavon}, {Stover}, {Tucker}, {Wallace}, {Wei}, {Wirth}, \&
  {Wright}}]{deimos}
{Faber}, S.~M., {Phillips}, A.~C., {Kibrick}, R.~I., {et~al.} 2003, in Society
  of Photo-Optical Instrumentation Engineers (SPIE) Conference Series, Vol.
  4841, Society of Photo-Optical Instrumentation Engineers (SPIE) Conference
  Series, ed. M.~{Iye} \& A.~F.~M. {Moorwood}, 1657--1669

\bibitem[{{Fioc} \& {Rocca-Volmerange}(1997)}]{fioc97}
{Fioc}, M. \& {Rocca-Volmerange}, B. 1997, \aap, 326, 950

\bibitem[{{Franceschini} {et~al.}(2008){Franceschini}, {Rodighiero}, \&
  {Vaccari}}]{franc08}
{Franceschini}, A., {Rodighiero}, G., \& {Vaccari}, M. 2008, \aap, 487, 837

\bibitem[{{Frayer} {et~al.}(2006){Frayer}, {Fadda}, {Yan}, {Marleau}, {Choi},
  {Helou}, {Soifer}, {Appleton}, {Armus}, {Beck}, {Dole}, {Engelbracht},
  {Fang}, {Gordon}, {Heinrichsen}, {Henderson}, {Hesselroth}, {Im}, {Kelly},
  {Lacy}, {Laine}, {Latter}, {Mahoney}, {Makovoz}, {Masci}, {Morrison},
  {Moshir}, {Noriega-Crespo}, {Padgett}, {Pesenson}, {Shupe}, {Squires},
  {Storrie-Lombardi}, {Surace}, {Teplitz}, \& {Wilson}}]{frayer}
{Frayer}, D.~T., {Fadda}, D., {Yan}, L., {et~al.} 2006, \aj, 131, 250

\bibitem[{{Gaztanaga}(1994)}]{gaztanaga}
{Gaztanaga}, E. 1994, Monthly Notices of the RAS, 268, 913

\bibitem[{{Genzel} \& {Cesarsky}(2000)}]{genzel00}
{Genzel}, R. \& {Cesarsky}, C.~J. 2000, \araa, 38, 761

\bibitem[{{Gilli} \& {Daddi}(2007)}]{gilli}
{Gilli}, R. \& {Daddi}, E. 2007, in ASP Conference Series, Vol. 380, Deepest
  Astronomical Surveys, ed. J.~{Afonso}, H.~C. {Ferguson}, B.~{Mobasher}, \&
  R.~{Norris}, 409

\bibitem[{{Goto}(2005)}]{goto05}
{Goto}, T. 2005, \mnras, 360, 322

\bibitem[{{Granato} {et~al.}(2000){Granato}, {Lacey}, {Silva}, {Bressan},
  {Baugh}, {Cole}, \& {Frenk}}]{granato}
{Granato}, G.~L., {Lacey}, C.~G., {Silva}, L., {et~al.} 2000, \apj, 542, 710

\bibitem[{{Guglielmo} {et~al.}(2015){Guglielmo}, {Poggianti}, {Moretti},
  {Fritz}, {Calvi}, {Vulcani}, {Fasano}, \& {Paccagnella}}]{guglielmo}
{Guglielmo}, V., {Poggianti}, B.~M., {Moretti}, A., {et~al.} 2015, \mnras, 450,
  2749

\bibitem[{{Gunn} \& {Gott}(1972)}]{gunn}
{Gunn}, J.~E. \& {Gott}, III, J.~R. 1972, \apj, 176, 1

\bibitem[{{Guzzo} {et~al.}(1997){Guzzo}, {Strauss}, {Fisher}, {Giovanelli}, \&
  {Haynes}}]{guzzo97}
{Guzzo}, L., {Strauss}, M.~A., {Fisher}, K.~B., {Giovanelli}, R., \& {Haynes},
  M.~P. 1997, \apj, 489, 37

\bibitem[{{Hauser} {et~al.}(1998){Hauser}, {Arendt}, {Kelsall}, {Dwek},
  {Odegard}, {Weiland}, {Freudenreich}, {Reach}, {Silverberg}, {Moseley},
  {Pei}, {Lubin}, {Mather}, {Shafer}, {Smoot}, {Weiss}, {Wilkinson}, \&
  {Wright}}]{hauserrr}
{Hauser}, M.~G., {Arendt}, R.~G., {Kelsall}, T., {et~al.} 1998, \apj, 508, 25

\bibitem[{{Heinis} {et~al.}(2009){Heinis}, {Budav{\'a}ri}, {Szalay}, {Arnouts},
  {Arag{\'o}n-Calvo}, {Wyder}, {Barlow}, {Foster}, {Peter}, {Martin},
  {Morrissey}, {Neff}, {Schiminovich}, {Seibert}, {Bianchi}, {Donas},
  {Heckman}, {Lee}, {Madore}, {Milliard}, {Rich}, \& {Yi}}]{heinis}
{Heinis}, S., {Budav{\'a}ri}, T., {Szalay}, A.~S., {et~al.} 2009, \apj, 698,
  1838

\bibitem[{{Hirschmann} {et~al.}(2013){Hirschmann}, {De Lucia}, {Iovino}, \&
  {Cucciati}}]{hirschmann}
{Hirschmann}, M., {De Lucia}, G., {Iovino}, A., \& {Cucciati}, O. 2013, \mnras,
  433, 1479

\bibitem[{{Hogg} {et~al.}(2004){Hogg}, {Blanton}, {Brinchmann}, {Eisenstein},
  {Schlegel}, {Gunn}, {McKay}, {Rix}, {Bahcall}, {Brinkmann}, \&
  {Meiksin}}]{hogg04}
{Hogg}, D.~W., {Blanton}, M.~R., {Brinchmann}, J., {et~al.} 2004, \apjl, 601,
  L29

\bibitem[{{Hopkins} {et~al.}(2001){Hopkins}, {Connolly}, \& {Szalay}}]{hop01}
{Hopkins}, A.~M., {Connolly}, A.~J., \& {Szalay}, A.~S. 2001, in Astronomical
  Society of the Pacific Conference Series, Vol. 240, Gas and Galaxy Evolution,
  ed. J.~E. {Hibbard}, M.~{Rupen}, \& J.~H. {van Gorkom}, 127

\bibitem[{{Houck} {et~al.}(2005){Houck}, {Soifer}, {Weedman}, {Higdon},
  {Higdon}, {Herter}, {Brown}, {Dey}, {Jannuzi}, {Le Floc'h}, {Rieke}, {Armus},
  {Charmandaris}, {Brandl}, \& {Teplitz}}]{houck}
{Houck}, J.~R., {Soifer}, B.~T., {Weedman}, D., {et~al.} 2005, \apjl, 622, L105

\bibitem[{Hsu {et~al.}(2003)Hsu, Chang, \& Lin}]{hsu}
Hsu, C.-W., Chang, C.-C., \& Lin, C.-J. 2003, Bioinformatics, 1, 1

\bibitem[{{Huertas-Company} {et~al.}(2008){Huertas-Company}, {Rouan}, {Tasca},
  {Soucail}, \& {Le F{\`e}vre}}]{hc}
{Huertas-Company}, M., {Rouan}, D., {Tasca}, L., {Soucail}, G., \& {Le
  F{\`e}vre}, O. 2008, \aap, 478, 971

\bibitem[{{Ilbert} {et~al.}(2006){Ilbert}, {Arnouts}, {McCracken},
  {Bolzonella}, {Bertin}, {Le F{\`e}vre}, {Mellier}, {Zamorani}, {Pell{\`o}},
  {Iovino}, {Tresse}, {Le Brun}, {Bottini}, {Garilli}, {Maccagni}, {Picat},
  {Scaramella}, {Scodeggio}, {Vettolani}, {Zanichelli}, {Adami}, {Bardelli},
  {Cappi}, {Charlot}, {Ciliegi}, {Contini}, {Cucciati}, {Foucaud}, {Franzetti},
  {Gavignaud}, {Guzzo}, {Marano}, {Marinoni}, {Mazure}, {Meneux}, {Merighi},
  {Paltani}, {Pollo}, {Pozzetti}, {Radovich}, {Zucca}, {Bondi}, {Bongiorno},
  {Busarello}, {de La Torre}, {Gregorini}, {Lamareille}, {Mathez}, {Merluzzi},
  {Ripepi}, {Rizzo}, \& {Vergani}}]{lephare}
{Ilbert}, O., {Arnouts}, S., {McCracken}, H.~J., {et~al.} 2006, \aap, 457, 841

\bibitem[{{Ilbert} {et~al.}(2009){Ilbert}, {Capak}, {Salvato}, {Aussel},
  {McCracken}, {Sanders}, {Scoville}, {Kartaltepe}, {Arnouts}, {Le Floc'h},
  {Mobasher}, {Taniguchi}, {Lamareille}, {Leauthaud}, {Sasaki}, {Thompson},
  {Zamojski}, {Zamorani}, {Bardelli}, {Bolzonella}, {Bongiorno}, {Brusa},
  {Caputi}, {Carollo}, {Contini}, {Cook}, {Coppa}, {Cucciati}, {de la Torre},
  {de Ravel}, {Franzetti}, {Garilli}, {Hasinger}, {Iovino}, {Kampczyk},
  {Kneib}, {Knobel}, {Kovac}, {Le Borgne}, {Le Brun}, {F{\`e}vre}, {Lilly},
  {Looper}, {Maier}, {Mainieri}, {Mellier}, {Mignoli}, {Murayama}, {Pell{\`o}},
  {Peng}, {P{\'e}rez-Montero}, {Renzini}, {Ricciardelli}, {Schiminovich},
  {Scodeggio}, {Shioya}, {Silverman}, {Surace}, {Tanaka}, {Tasca}, {Tresse},
  {Vergani}, \& {Zucca}}]{smetnydred}
{Ilbert}, O., {Capak}, P., {Salvato}, M., {et~al.} 2009, \apj, 690, 1236

\bibitem[{{Kaiser}(1984)}]{kaiser}
{Kaiser}, N. 1984, \apjl, 284, L9

\bibitem[{{Lagache} {et~al.}(2007){Lagache}, {Bavouzet}, {Fernandez-Conde},
  {Ponthieu}, {Rodet}, {Dole}, {Miville-Desch{\^e}nes}, \& {Puget}}]{lagache07}
{Lagache}, G., {Bavouzet}, N., {Fernandez-Conde}, N., {et~al.} 2007, \apjl,
  665, L89

\bibitem[{{Lagache} {et~al.}(2004){Lagache}, {Dole}, {Puget},
  {P{\'e}rez-Gonz{\'a}lez}, {Le Floc'h}, {Rieke}, {Papovich}, {Egami},
  {Alonso-Herrero}, {Engelbracht}, {Gordon}, {Misselt}, \&
  {Morrison}}]{lagache}
{Lagache}, G., {Dole}, H., {Puget}, J.-L., {et~al.} 2004, \apjs, 154, 112

\bibitem[{{Lagache} {et~al.}(2005){Lagache}, {Puget}, \& {Dole}}]{lagache05}
{Lagache}, G., {Puget}, J.-L., \& {Dole}, H. 2005, \araa, 43, 727

\bibitem[{{Landy} \& {Szalay}(1993)}]{ls}
{Landy}, S.~D. \& {Szalay}, A.~S. 1993, \apj, 412, 64

\bibitem[{{Le Floc'h} {et~al.}(2005){Le Floc'h}, {Papovich}, {Dole}, {Bell},
  {Lagache}, {Rieke}, {Egami}, {P{\'e}rez-Gonz{\'a}lez}, {Alonso-Herrero},
  {Rieke}, {Blaylock}, {Engelbracht}, {Gordon}, {Hines}, {Misselt}, {Morrison},
  \& {Mould}}]{le05}
{Le Floc'h}, E., {Papovich}, C., {Dole}, H., {et~al.} 2005, \apj, 632, 169

\bibitem[{{Limber}(1953)}]{limb}
{Limber}, D.~N. 1953, \apj, 117, 134

\bibitem[{{Lonsdale} \& {Hacking}(1989)}]{iras}
{Lonsdale}, C.~J. \& {Hacking}, P.~B. 1989, \apj, 339, 712

\bibitem[{{Magliocchetti} {et~al.}(2008){Magliocchetti}, {Cirasuolo}, {McLure},
  {Dunlop}, {Almaini}, {Foucaud}, {de Zotti}, {Simpson}, \&
  {Sekiguchi}}]{maglio}
{Magliocchetti}, M., {Cirasuolo}, M., {McLure}, R.~J., {et~al.} 2008, \mnras,
  383, 1131

\bibitem[{{Ma{\l}ek} {et~al.}(2014){Ma{\l}ek}, {Pollo}, {Takeuchi}, {Buat},
  {Burgarella}, {Malkan}, {Giovannoli}, {Kurek}, \& {Matsuura}}]{kaska14}
{Ma{\l}ek}, K., {Pollo}, A., {Takeuchi}, T.~T., {et~al.} 2014, \aap, 562, A15

\bibitem[{{Malek} {et~al.}(2012){Malek}, {Pollo}, {Takeuchi}, {Giovannoli},
  {Buat}, {Burgarella}, \& {Malkan}}]{malek12}
{Malek}, K., {Pollo}, A., {Takeuchi}, T.~T., {et~al.} 2012, Publication of
  Korean Astronomical Society, 27, 141

\bibitem[{{Ma{\l}ek} {et~al.}(2013){Ma{\l}ek}, {Solarz}, {Pollo}, {Fritz},
  {Garilli}, {Scodeggio}, {Iovino}, {Granett}, {Abbas}, {Adami}, {Arnouts},
  {Bel}, {Bolzonella}, {Bottini}, {Branchini}, {Cappi}, {Coupon}, {Cucciati},
  {Davidzon}, {De Lucia}, {de la Torre}, {Franzetti}, {Fumana}, {Guzzo},
  {Ilbert}, {Krywult}, {Le Brun}, {Le Fevre}, {Maccagni}, {Marulli},
  {McCracken}, {Paioro}, {Polletta}, {Schlagenhaufer}, {Tasca}, {Tojeiro},
  {Vergani}, {Zanichelli}, {Burden}, {Di Porto}, {Marchetti}, {Marinoni},
  {Mellier}, {Moscardini}, {Nichol}, {Peacock}, {Percival}, {Phleps}, {Wolk},
  \& {Zamorani}}]{my}
{Ma{\l}ek}, K., {Solarz}, A., {Pollo}, A., {et~al.} 2013, \aap, 557, A16

\bibitem[{{Maraston}(2005)}]{maraston05}
{Maraston}, C. 2005, \mnras, 362, 799

\bibitem[{{Marinoni} {et~al.}(2002){Marinoni}, {Hudson}, \&
  {Giuricin}}]{marinoni02}
{Marinoni}, C., {Hudson}, M.~J., \& {Giuricin}, G. 2002, \apj, 569, 91

\bibitem[{{Marinoni} {et~al.}(1999){Marinoni}, {Monaco}, {Giuricin}, \&
  {Costantini}}]{marinoni99}
{Marinoni}, C., {Monaco}, P., {Giuricin}, G., \& {Costantini}, B. 1999, \apj,
  521, 50

\bibitem[{{Masci} \& {SWIRE Team}(2006)}]{masci}
{Masci}, F.~J. \& {SWIRE Team}. 2006, in ASP Conference Series, Vol. 357, ASP
  Conference Series, ed. L.~{Armus} \& W.~T. {Reach}, 271

\bibitem[{{Matsuhara} {et~al.}(2006{\natexlab{a}}){Matsuhara}, {Wada},
  {Matsuura}, {Nakagawa}, {Kawada}, {Ohyama}, {Pearson}, {Oyabu}, {Takagi},
  {Serjeant}, {White}, {Hanami}, {Watarai}, {Takeuchi}, {Kodama}, {Arimoto},
  {Okamura}, {Lee}, {Pak}, {Im}, {Lee}, {Kim}, {Jeong}, {Imai}, {Fujishiro},
  {Shirahata}, {Suzuki}, {Ihara}, \& {Sakon}}]{matsu06}
{Matsuhara}, H., {Wada}, T., {Matsuura}, S., {et~al.} 2006{\natexlab{a}},
  \pasj, 58, 673

\bibitem[{{Matsuhara} {et~al.}(2006{\natexlab{b}}){Matsuhara}, {Wada},
  {Matsuura}, {Nakagawa}, {Kawada}, {Ohyama}, {Pearson}, {Oyabu}, {Takagi},
  {Serjeant}, {White}, {Hanami}, {Watarai}, {Takeuchi}, {Kodama}, {Arimoto},
  {Okamura}, {Lee}, {Pak}, {Im}, {Lee}, {Kim}, {Jeong}, {Imai}, {Fujishiro},
  {Shirahata}, {Suzuki}, {Ihara}, \& {Sakon}}]{matsuhara}
{Matsuhara}, H., {Wada}, T., {Matsuura}, S., {et~al.} 2006{\natexlab{b}},
  \pasj, 58, 673

\bibitem[{{Melbourne} {et~al.}(2005){Melbourne}, {Koo}, \& {Le
  Floc'h}}]{melbourne05}
{Melbourne}, J., {Koo}, D.~C., \& {Le Floc'h}, E. 2005, \apjl, 632, L65

\bibitem[{{Meneux} {et~al.}(2006){Meneux}, {Le F{\`e}vre}, {Guzzo}, {Pollo},
  {Cappi}, {Ilbert}, {Iovino}, {Marinoni}, {McCracken}, {Bottini}, {Garilli},
  {Le Brun}, {Maccagni}, {Picat}, {Scaramella}, {Scodeggio}, {Tresse},
  {Vettolani}, {Zanichelli}, {Adami}, {Arnouts}, {Arnaboldi}, {Bardelli},
  {Bolzonella}, {Charlot}, {Ciliegi}, {Contini}, {Foucaud}, {Franzetti},
  {Gavignaud}, {Marano}, {Mazure}, {Merighi}, {Paltani}, {Pell{\`o}},
  {Pozzetti}, {Radovich}, {Zamorani}, {Zucca}, {Bondi}, {Bongiorno},
  {Busarello}, {Cucciati}, {Gregorini}, {Lamareille}, {Mathez}, {Mellier},
  {Merluzzi}, {Ripepi}, \& {Rizzo}}]{meneux06}
{Meneux}, B., {Le F{\`e}vre}, O., {Guzzo}, L., {et~al.} 2006, \aap, 452, 387

\bibitem[{{Milliard} {et~al.}(2007){Milliard}, {Heinis}, {Blaizot}, {Arnouts},
  {Schiminovich}, {Budav{\'a}ri}, {Donas}, {Treyer}, {Laget}, {Viton}, {Wyder},
  {Szalay}, {Barlow}, {Forster}, {Friedman}, {Martin}, {Morrissey}, {Neff},
  {Seibert}, {Small}, {Bianchi}, {Heckman}, {Lee}, {Madore}, {Rich}, {Welsh},
  {Yi}, \& {Xu}}]{milliard}
{Milliard}, B., {Heinis}, S., {Blaizot}, J., {et~al.} 2007, \apjs, 173, 494

\bibitem[{{Mo} \& {White}(1996)}]{mowhite}
{Mo}, H.~J. \& {White}, S.~D.~M. 1996, \mnras, 282, 347

\bibitem[{{Moore} {et~al.}(1996){Moore}, {Katz}, {Lake}, {Dressler}, \&
  {Oemler}}]{moore}
{Moore}, B., {Katz}, N., {Lake}, G., {Dressler}, A., \& {Oemler}, A. 1996,
  \nat, 379, 613

\bibitem[{{Moscardini} {et~al.}(1998){Moscardini}, {Coles}, {Lucchin}, \&
  {Matarrese}}]{mosc98}
{Moscardini}, L., {Coles}, P., {Lucchin}, F., \& {Matarrese}, S. 1998, \mnras,
  299, 95

\bibitem[{{Murata} {et~al.}(2013){Murata}, {Matsuhara}, {Wada}, {Arimatsu},
  {Oi}, {Takagi}, {Oyabu}, {Goto}, {Ohyama}, {Malkan}, {Pearson}, {Ma{\l}ek},
  \& {Solarz}}]{murata13}
{Murata}, K., {Matsuhara}, H., {Wada}, T., {et~al.} 2013, \aap, 559, A132

\bibitem[{{Noll} {et~al.}(2009){Noll}, {Burgarella}, {Giovannoli},
  {et~al.}}]{noll09}
{Noll}, S., {Burgarella}, D., {Giovannoli}, E., {et~al.} 2009, A\&A, 507, 1793

\bibitem[{{Nusser} \& {Davis}(1994)}]{nusser94}
{Nusser}, A. \& {Davis}, M. 1994, \apjl, 421, L1

\bibitem[{{Onaka} {et~al.}(2007){Onaka}, {Matsuhara}, {Wada}, {Fujishiro},
  {Fujiwara}, {Ishigaki}, {Ishihara}, {Ita}, {Kataza}, {Kim}, {Matsumoto},
  {Murakami}, {Ohyama}, {Oyabu}, {Sakon}, {Tanab{\'e}}, {Takagi}, {Uemizu},
  {Ueno}, {Usui}, {Watarai}, {Cohen}, {Enya}, {Ootsubo}, {Pearson}, {Takeyama},
  {Yamamuro}, \& {Ikeda}}]{onaka}
{Onaka}, T., {Matsuhara}, H., {Wada}, T., {et~al.} 2007, \pasj, 59, 401

\bibitem[{{Papovich} {et~al.}(2004){Papovich}, {Dole}, {Egami}, {Le Floc'h},
  {P{\'e}rez-Gonz{\'a}lez}, {Alonso-Herrero}, {Bai}, {Beichman}, {Blaylock},
  {Engelbracht}, {Gordon}, {Hines}, {Misselt}, {Morrison}, {Mould},
  {Muzerolle}, {Neugebauer}, {Richards}, {Rieke}, {Rieke}, {Rigby}, {Su}, \&
  {Young}}]{spitzer}
{Papovich}, C., {Dole}, H., {Egami}, E., {et~al.} 2004, \apjs, 154, 70

\bibitem[{{Peebles}(1980)}]{pee80}
{Peebles}, P.~J.~E. 1980, {The large-scale structure of the universe}

\bibitem[{Peebles(1994)}]{pee}
Peebles, P.~J.~E. 1994, Physical Cosmology (Princeton, NJ: Princeton University
  Press)

\bibitem[{{Peebles} \& {Groth}(1976)}]{peegroth}
{Peebles}, P.~J.~E. \& {Groth}, E.~J. 1976, \aap, 53, 131

\bibitem[{{P{\'e}rez-Gonz{\'a}lez} {et~al.}(2005){P{\'e}rez-Gonz{\'a}lez},
  {Rieke}, {Egami}, {Alonso-Herrero}, {Dole}, {Papovich}, {Blaylock}, {Jones},
  {Rieke}, {Rigby}, {Barmby}, {Fazio}, {Huang}, \& {Martin}}]{perez-gonzalez05}
{P{\'e}rez-Gonz{\'a}lez}, P.~G., {Rieke}, G.~H., {Egami}, E., {et~al.} 2005,
  \apj, 630, 82

\bibitem[{{Planck Collaboration} {et~al.}(2014){Planck Collaboration}, {Ade},
  {Aghanim}, {Alves}, {Armitage-Caplan}, {Arnaud}, {Ashdown},
  {Atrio-Barandela}, {Aumont}, {Aussel}, \& et~al.}]{planck1}
{Planck Collaboration}, {Ade}, P.~A.~R., {Aghanim}, N., {et~al.} 2014, \aap,
  571, A1

\bibitem[{{Pollo} {et~al.}(2005){Pollo}, {Meneux}, {Guzzo}, {Le F{\`e}vre},
  {Blaizot}, {Cappi}, {Iovino}, {Marinoni}, {McCracken}, {Bottini}, {Garilli},
  {Le Brun}, {Maccagni}, {Picat}, {Scaramella}, {Scodeggio}, {Tresse},
  {Vettolani}, {Zanichelli}, {Adami}, {Arnaboldi}, {Arnouts}, {Bardelli},
  {Bolzonella}, {Charlot}, {Ciliegi}, {Contini}, {Foucaud}, {Franzetti},
  {Gavignaud}, {Ilbert}, {Marano}, {Mathez}, {Mazure}, {Merighi}, {Paltani},
  {Pell{\`o}}, {Pozzetti}, {Radovich}, {Zamorani}, {Zucca}, {Bondi},
  {Bongiorno}, {Busarello}, {Gregorini}, {Lamareille}, {Mellier}, {Merluzzi},
  {Ripepi}, \& {Rizzo}}]{pollo05}
{Pollo}, A., {Meneux}, B., {Guzzo}, L., {et~al.} 2005, A\&A, 439, 887

\bibitem[{{Quadri}(2007)}]{quadri}
{Quadri}, R.~F. 2007, PhD thesis, Yale University

\bibitem[{{Roche} \& {Eales}(1999)}]{roche}
{Roche}, N. \& {Eales}, S.~A. 1999, \mnras, 307, 703

\bibitem[{{Ross} {et~al.}(2007){Ross}, {da {\^A}ngela}, {Shanks}, {Wake},
  {Cannon}, {Edge}, {Nichol}, {Outram}, {Colless}, {Couch}, {Croom}, {de
  Propris}, {Drinkwater}, {Eisenstein}, {Loveday}, {Pimbblet}, {Roseboom},
  {Schneider}, {Sharp}, \& {Weilbacher}}]{ross}
{Ross}, N.~P., {da {\^A}ngela}, J., {Shanks}, T., {et~al.} 2007, Monthly
  Notices of the RAS, 381, 573

\bibitem[{{Scoccimarro} {et~al.}(2001){Scoccimarro}, {Sheth}, {Hui}, \&
  {Jain}}]{hod}
{Scoccimarro}, R., {Sheth}, R.~K., {Hui}, L., \& {Jain}, B. 2001, \apj, 546, 20

\bibitem[{{Scranton} {et~al.}(2002){Scranton}, {Johnston}, {Dodelson},
  {Frieman}, {Connolly}, {Eisenstein}, {Gunn}, {Hui}, {Jain}, {Kent},
  {Loveday}, {Narayanan}, {Nichol}, {O'Connell}, {Scoccimarro}, {Sheth},
  {Stebbins}, {Strauss}, {Szalay}, {Szapudi}, {Tegmark}, {Vogeley}, {Zehavi},
  {Annis}, {Bahcall}, {Brinkman}, {Csabai}, {Hindsley}, {Ivezic}, {Kim},
  {Knapp}, {Lamb}, {Lee}, {Lupton}, {McKay}, {Munn}, {Peoples}, {Pier},
  {Richards}, {Rockosi}, {Schlegel}, {Schneider}, {Stoughton}, {Tucker},
  {Yanny}, \& {York}}]{scrantan}
{Scranton}, R., {Johnston}, D., {Dodelson}, S., {et~al.} 2002, \apj, 579, 48

\bibitem[{{Shawe-Taylor} \& {Cristianini}(2004)}]{st}
{Shawe-Taylor}, S. \& {Cristianini}, N. 2004, Kernel Methods for Pattern
  Analysis (Cambridge, UK: Cambridge, UP)

\bibitem[{{Sheth} \& {Tormen}(1999)}]{sheth}
{Sheth}, R.~K. \& {Tormen}, G. 1999, \mnras, 308, 119

\bibitem[{{Solarz} {et~al.}(2012){Solarz}, {Pollo}, {Takeuchi}, {P{\c e}piak},
  {Matsuhara}, {Wada}, {Oyabu}, {Takagi}, {Goto}, {Ohyama}, {Pearson},
  {Hanami}, \& {Ishigaki}}]{ja}
{Solarz}, A., {Pollo}, A., {Takeuchi}, T.~T., {et~al.} 2012, \aap, 541, A50

\bibitem[{{Starikova} {et~al.}(2012){Starikova}, {Berta}, {Franceschini},
  {Marchetti}, {Rodighiero}, {Vaccari}, \& {Vikhlinin}}]{starikova}
{Starikova}, S., {Berta}, S., {Franceschini}, A., {et~al.} 2012, \apj, 751, 126

\bibitem[{{Sullivan} {et~al.}(2001){Sullivan}, {Mobasher}, {Chan}, {Cram},
  {Ellis}, {Treyer}, \& {Hopkins}}]{sullivan01}
{Sullivan}, M., {Mobasher}, B., {Chan}, B., {et~al.} 2001, \apj, 558, 72

\bibitem[{{Takeuchi} {et~al.}(2005){Takeuchi}, {Buat}, \&
  {Burgarella}}]{takeuchi05}
{Takeuchi}, T.~T., {Buat}, V., \& {Burgarella}, D. 2005, \aap, 440, L17

\bibitem[{{Tanab{\'e}} {et~al.}(2008){Tanab{\'e}}, {Sakon}, {Cohen}, {Wada},
  {Ita}, {Ohyama}, {Oyabu}, {Uemizu}, {Takagi}, {Ishihara}, {Kim}, {Ueno},
  {Matsuhara}, \& {Onaka}}]{tan}
{Tanab{\'e}}, T., {Sakon}, I., {Cohen}, M., {et~al.} 2008, \pasj, 60, 375

\bibitem[{{Taniguchi} {et~al.}(1998){Taniguchi}, {Trentham}, \&
  {Shioya}}]{taniguchi}
{Taniguchi}, Y., {Trentham}, N., \& {Shioya}, Y. 1998, \apjl, 504, L79

\bibitem[{{Toomre} \& {Toomre}(1972)}]{toomre}
{Toomre}, A. \& {Toomre}, J. 1972, \apj, 178, 623

\bibitem[{{Wada} {et~al.}(2008){Wada}, {Matsuhara}, {Oyabu}, {Takagi}, {Lee},
  {Im}, {Ohyama}, {Goto}, {Pearson}, {White}, {Serjeant}, {Wada}, \&
  {Hanami}}]{wada}
{Wada}, T., {Matsuhara}, H., {Oyabu}, S., {et~al.} 2008, \pasj, 60, 517

\bibitem[{{Weinmann} {et~al.}(2006){Weinmann}, {van den Bosch}, {Yang}, \&
  {Mo}}]{weinmann06}
{Weinmann}, S.~M., {van den Bosch}, F.~C., {Yang}, X., \& {Mo}, H.~J. 2006,
  \mnras, 366, 2

\bibitem[{{Wo{\'z}niak} {et~al.}(2004){Wo{\'z}niak}, {Williams}, {Vestrand}, \&
  {Gupta}}]{wozniak}
{Wo{\'z}niak}, P.~R., {Williams}, S.~J., {Vestrand}, W.~T., \& {Gupta}, V.
  2004, \aj, 128, 2965

\bibitem[{{Zhang} \& {Zhao}(2004)}]{zz}
{Zhang}, Y. \& {Zhao}, Y. 2004, \aap, 422, 1113

\end{thebibliography}
\end{document}